\documentclass[3p,review]{elsarticle}
\usepackage{hyperref}

\usepackage{float}
\usepackage{subfigure}
\usepackage[export]{adjustbox}
\usepackage{url}
\usepackage{caption}

\usepackage{array}

\usepackage{tikz}
\usetikzlibrary{shapes.geometric}

\usepackage{amsmath,bm}
\usepackage{multicol,multirow,lipsum}
\usepackage{romannum}

\usepackage{xcolor}
\definecolor{denim}{rgb}{0.08, 0.38, 0.74}
\definecolor{almond}{rgb}{0.94, 0.87, 0.8}
\definecolor{asparagus}{rgb}{0.53, 0.66, 0.42}
\definecolor{antiquebrass}{rgb}{0.8, 0.58, 0.46}
\definecolor{bluebell}{rgb}{0.64, 0.64, 0.82}
\definecolor{cadetblue}{rgb}{0.37, 0.62, 0.63}
\definecolor{jade}{rgb}{0.0, 0.66, 0.42}
\definecolor{blush}{rgb}{0.87, 0.36, 0.51}
\definecolor{amethyst}{rgb}{0.6, 0.4, 0.8}
\definecolor{bostonuniversityred}{rgb}{0.8, 0.0, 0.0}
\definecolor{brandeisblue}{rgb}{0.0, 0.44, 1.0}
\definecolor{applegreen}{rgb}{0.55, 0.71, 0.0}

\definecolor{mediumseagreen}{rgb}{0.24, 0.7, 0.44}
\definecolor{orchid}{rgb}{0.85, 0.44, 0.84}
\definecolor{dodgerblue}{rgb}{0.12, 0.56, 1.0}
\definecolor{mediumpurple}{rgb}{0.58, 0.44, 0.86}
\definecolor{sandybrown}{rgb}{0.96, 0.64, 0.38}
\definecolor{olive}{rgb}{0.5, 0.5, 0.0}
\definecolor{sienna}{rgb}{0.53, 0.18, 0.09}

\begin{document}

\pagenumbering{arabic} 
\begin{frontmatter}

\title{Physics-informed neural networks incorporating energy dissipation 
 for the phase-field model of ferroelectric microstructure evolution}

\author[1]{Lan Shang}
\ead{shanglan@zhejianglab.com}
\author[2]{Sizheng Zheng}
\author[2]{Jin Wang}
\author[1,2]{Jie Wang}
\ead{jw@zju.edu.cn}
\address[1]{Zhejiang Lab, Hangzhou, China}
\address[2]{Department of Engineering Mechanics, Zhejiang University, Hangzhou, China}

\begin{abstract}
        {Physics-informed neural networks (PINNs) are an emerging technique 
         to solve partial differential equations (PDEs). In this work, we 
         propose a simple but effective PINN approach for the phase-field model of 
         ferroelectric microstructure evolution. This model is a time-dependent, nonlinear, 
         and high-order PDE system of multi-physics, challenging to be solved 
         using a baseline PINN. Considering that the acquisition of steady 
         microstructures is one of the primary focuses in simulations of 
         ferroelectric microstructure evolution, we simplify the time-dependent 
         PDE system to be a static problem. This static problem, however, 
         is ill-posed. To overcome this issue, a term originated from the law of 
         energy dissipation is embedded into the loss function as an extra 
         constraint for the PINN. With this modification, the PINN successfully   
         predicts the steady ferroelectric microstructure without tracking the 
         evolution process. In addition, although the proposed PINN approach cannot 
         tackle the dynamic problem in a straightforward fashion, it is of benefit to 
         the PINN prediction of the evolution process by providing labeled data. 
         These data are crucial because they help the PINN avoid the propagation failure, 
         a common failure mode of PINNs when predicting dynamic behaviors.
         The above mentioned advantages of the proposed PINN approach are demonstrated 
         through a number of examples.
         }  
\end{abstract} 
\begin{keyword}
        physics-informed neural network 
        \sep
        phase-field model
        \sep
        ferroelectric microstructure evolution  
        \sep
        propagation failure 
        \sep
        energy dissipation
        \sep
        loss function  
\end{keyword}
\end{frontmatter}

\section{Introduction} \label{sect_introduction}
Physics-informed neural networks (PINNs) \cite{dissanayake1994neural, lagaris1998artificial, raissi2019physics} 
are an alternative method \cite{markidis2021old} to solve partial differential equations (PDEs). 
They have recently become a vibrant topic in scientific machine learning 
\cite{karniadakis2021physics, cuomo2022scientific} with applications 
across many fields \cite{raissi2020hidden, yazdani2020systems, haghighat2021physics, 
huang2022applications, diao2023solving, jiang2023practical}. 

Despite the empirical success of the baseline PINN \cite{raissi2019physics} in a variety of PDEs, 
it is also well-known that it struggles or even fails to solve some problems 
\cite{krishnapriyan2021characterizing, rohrhofer2022role, wang2022and, chuang2023predictive, bajaj2023recipes, sharma2023stiff}. 
Since neural networks have great expressivity \cite{scarselli1998universal}, 
the difficulties are from the optimization procedure,    
i.e., the training in machine learning. 
A PINN is trained to minimize a loss function derived from the PDE problem with 
a few or without labeled data, so the loss landscape is highly dependent on the 
PDE characteristics, causing various issues confronted by the optimization, 
e.g., high computational costs \cite{grossmann2024can}, 
unbalanced gradients \cite{wang2021understanding}, 
spectral bias \cite{rahaman2019spectral}, and ill-conditioning \cite{rathore2024challenges}. 
A range of modifications to the baseline PINN has been reported to address these issues, 
including but not limited to domain decomposition \cite{jagtap2020extended, shukla2021parallel}, 
dynamic loss weighting \cite{deguchi2023dynamic, zhang2024physics}, 
adaptive sampling \cite{gao2023failure, wu2023comprehensive}, 
causal training \cite{mattey2022novel, wang2022respecting, penwarden2023unified}, 
Fourier feature embeddings \cite{wang2021eigenvector}, 
adaptive activation functions \cite{jagtap2020adaptive, jagtap2022deep}, 
weak form based loss functions \cite{Kharazmi2019Variational, samaniego2020energy}, 
and enhanced optimizers \cite{rathore2024challenges, kopanivcakova2024enhancing}. 
All the methods can make the PINN training easier for 
some PDEs, but many problems remain challenging when using PINNs, 
so strategies with further improvements are still to be explored.

This work proposes a new PINN approach to solve the phase-field model 
of ferroelectric microstructure evolution. Study on microstructure
evolution plays an essential role in understanding and developing ferroelectric
materials \cite{nelson2011domain}, which are a class of multifunctional materials 
widely used in actuators, sensors, and information storage due to their unique 
electromechanical properties. To simulate ferroelectric microstructure evolution, 
one common method is to solve a PDE system comprising the 
time-dependent Ginzburg-Landau (TDGL) equation, the equation of momentum balance, 
and the Maxwell's equation, as well as appropriate boundary and initial conditions. 
The TDGL equation, also known as the Allen-Cahn equation, together with the 
Cahn-Hilliard equation constitutes the primary equations in the phase-field method 
\cite{moelans2008introduction, chen2022classical}, so the above PDE system is called 
a phase-field model. This model is high-order and nonlinear, normally 
solved using the Fourier-spectral method 
\cite{chen1998applications, durdiev2023effective} 
or the finite element method (FEM) \cite{dong2012finite, fedeli2019phase}. 
It also makes sense to solve phase-field models with PINNs, and relevant investigations 
can be found in \cite{mattey2022novel, wight2020solving, li2023phase, mattey2024gradient, qiumei2024mass}. 
However, these papers \cite{mattey2022novel, wight2020solving, li2023phase, mattey2024gradient, qiumei2024mass} 
concern only a single Allen-Cahn or Cahn-Hilliard equation, while the phase-field model 
of interest in the current work involves a group of coupled PDEs. 
Our previous work \cite{shang2024quantification} 
attempts to solve this phase-field model using PINNs. 
We find that it takes great effort to obtain the steady 
ferroelectric microstructure from a specified initial condition. 
To reach this goal, multiple independent PINNs are sequentially trained in 
\cite{shang2024quantification}, each for one time segment of the entire evolution process. 
In the current work, we propose a more straightforward PINN approach by taking 
advantage of extra physics beyond the PDE system to directly predict  
the steady microstructure of a ferroelectric material. 

The recent study \cite{qiumei2024mass} resorts to extra physics, too. 
It embeds mass conservation into the loss function of the PINN formulation 
for the Cahn-Hilliard equation to gain better prediction performance. 
The extra physics we utilize in this work is energy dissipation \cite{yang2017numerical}.
More specifically, the physical law that governs ferroelectric microstructure evolution states 
that the total free energy of a system decreases when the system evolves toward the equilibrium. 
Therefore, the system at equilibrium will have the smallest total free energy. 
Based on that, we can obtain the ferroelectric microstructure at equilibrium using a PINN 
without tracking the evolution process.
Furthermore, taking this microstructure information as labeled data, 
more results of interest, e.g., the evolution process, will be accessed more easily.

We note that the proposed method is not the deep energy method 
\cite{samaniego2020energy, nguyen2020deep}, 
which solves PDEs with PINNs by simply minimizing the total energy of the system.  
The deep energy method is useful only when the solution to the PDE is determined 
by the minimum energy state of the system, e.g., static linear-elastic problems \cite{samaniego2020energy}. 
Nevertheless, this is not true for the case considered in our work. In other words, 
a minimized total free energy of a ferroelectric material cannot guarantee 
the correct microstructure at equilibrium. To achieve a correct equilibrium state, 
our work actually minimizes the total free energy together with the residuals of a static PDE system. 
Hence, the loss function has a mixed formulation of the weak form and strong form. From this perspective, 
the proposed method is similar to papers \cite{fuhg2022mixed, rezaei2022mixed, harandi2024mixed}. 
The fundamental difference between our work and papers \cite{fuhg2022mixed, rezaei2022mixed, harandi2024mixed} 
consists in the motivation of adding the energy form to the loss function. 
In our work, the energy form is indispensable to avoid a trivial solution, 
while papers \cite{fuhg2022mixed, rezaei2022mixed, harandi2024mixed} 
introduce the energy form largely for higher computational efficiency. This difference 
eventually gives rise to distinct details of the loss functions between our work and 
papers \cite{fuhg2022mixed, rezaei2022mixed, harandi2024mixed}.

The remainder of this work is structured as follows. 
Section \ref{sect_overview_of_pinn} 
briefly reviews the basic idea of PINNs. 
Section \ref{sect_phase_field_model} 
presents the phase-field model of ferroelectric microstructure evolution. 
Section \ref{sect_energy_dissipation} 
proposes the method of incorporating energy dissipation into the PINN framework. 
Section \ref{sect_examples} 
validates the proposed method through several examples of static problems.
Section \ref{sect_dynamic_example} handles a dynamic problem, aimed at 
demonstrating that the proposed PINN approach in combination with the baseline 
PINN can produce much more accurate results.
Section \ref{sect_conclusions}
summarizes the main contributions of this work.

\section{Overview of physics-informed neural networks (PINNs)}
\label{sect_overview_of_pinn}
PINNs approximate the solution to a well-posed PDE problem in the form of 
a neural network using the mathematical expressions involved in the problem 
rather than many data. The key idea here is to construct a loss function 
that represents the residuals of the governing PDE and the boundary/initial conditions 
at an adequately large number of collocation points.
This loss function is then minimized through an optimization procedure to determine 
all tunable parameters of the network (e.g., weights and biases). 
Once done, taking a set of independent variables of the governing PDE as the network inputs, 
the outputs of the network are the corresponding solution. 

To illustrate the above, we consider a well-posed PDE problem as follows: 
\begin{subequations}
\begin{align}
       & \mathcal{F}[\bm{v}({\bm{x}},t)] = 0, \quad \bm{x} \in \Omega, \quad t \in [0, T], \label{eq_pde} \\ 
       & \mathcal{B}[\bm{v}({\bm{x}},t)] = 0, \quad \bm{x} \in \partial{\Omega}, \quad t \in [0, T], \label{eq_bc} \\
       & \bm{v}(\bm{x}, 0) = \bm{g}(\bm{x}), \quad \bm{x} \in \Omega \label{eq_ic}, 
\end{align}
\label{eq_PDE_problem}
\end{subequations}
where $\bm{x}$ and $t$ are the space-time coordinates, 
$\bm{v}$ is the latent solution,
$\mathcal{F}[\cdot]$ is a differential operator, 
$\mathcal{B}[\cdot]$ is a boundary operator, 
and $\bm{g}(\bm{x})$ is a known function. 
If using a neural network $\mathcal{N}(\bm{x}, t; \Theta)$ 
with tunable parameters $\Theta$ to approximate the solution $\bm{v}(\bm{x},t)$, 
i.e., $\mathcal{N}(\bm{x}, t; \Theta) = \bm{\tilde{v}}(\bm{x},t) \approx 
\bm{v}(\bm{x},t)$, the loss function can be written as 
\begin{align}
        \mathcal{L}(\Theta) = \lambda_{\text{pde}}\mathcal{L}_{\text{pde}} 
                             +\lambda_{\text{bc}}\mathcal{L}_{\text{bc}} 
                             +\lambda_{\text{ic}}\mathcal{L}_{\text{ic}}, 
\label{eq_total_loss}
\end{align}
where the coefficients $\lambda_{\text{pde}}, \lambda_{\text{bc}}, \lambda_{\text{ic}}$
are the weights for different terms in the loss function, 
allowing for better convergence of the PINN training 
if they are well assigned \cite{wang2021understanding}, 
and terms $\mathcal{L}_{\text{pde}}, \mathcal{L}_{\text{bc}}, 
\mathcal{L}_{\text{ic}}$ are usually computed by 
\begin{subequations}
\begin{align}
        &\mathcal{L}_{\text{pde}} = \frac{1}{N_{\text{pde}}}\sum_{i=1}^{N_{\text{pde}}}
         \bigg |\mathcal{F}[\bm{\tilde{v}}(\bm{x}_{\text{pde}}^{[i]}, t_{\text{pde}}^{[i]})] \bigg |^2
         := |\mathcal{F}[\bm{\tilde{v}}] |_{\Omega \times [0, T]}^2, 
         \label{eq_loss_pde} \\
        &\mathcal{L}_{\text{bc}} = \frac{1}{N_{\text{bc}}}\sum_{i=1}^{N_{\text{bc}}}
         \bigg |\mathcal{B}[\bm{\tilde{v}}(\bm{x}_{\text{bc}}^{[i]}, t_{\text{bc}}^{[i]})] \bigg |^2 
         := |\mathcal{B}[\bm{\tilde{v}}] |_{\partial{\Omega}\times [0, T]}^2,
         \label{eq_loss_bc} \\
        &\mathcal{L}_{\text{ic}} = \frac{1}{N_{\text{ic}}}\sum_{i=1}^{N_{\text{ic}}}
         \bigg |\bm{\tilde{v}}(\bm{x}_{\text{ic}}^{[i]}, 0) - \bm{g}(\bm{x}_{\text{ic}}^{[i]}) \bigg |^2 
         :=  |\bm{\tilde{v}} - \bm{g}(\bm{x}) |_{\Omega \times [t = 0]}^2.
         \label{eq_loss_ic} 
\end{align}
\label{eq_loss_components}
\end{subequations}
In Eq. (\ref{eq_loss_components}), $\{(\bm{x}_{\text{pde}}^{[i]}, t_{\text{pde}}^{[i]})\}_{i = 1}^{N_{\text{pde}}}$, 
$\{(\bm{x}_{\text{bc}}^{[i]}, t_{\text{bc}}^{[i]})\}_{i = 1}^{N_{\text{bc}}}$, 
and $\{(\bm{x}_{\text{ic}}^{[i]}, 0)\}_{i = 1}^{N_{\text{ic}}}$ are three sets of points 
sampled in the computational domain $\Omega \times [0, T]$  
to evaluate the losses associated with the governing PDE Eq. (\ref{eq_pde}), 
boundary condition Eq. (\ref{eq_bc}), 
and initial condition Eq. (\ref{eq_ic}), respectively. These points are collocation points, 
and how to sample them is an active field of research \cite{wu2023comprehensive}. 
In addition to soft constraints of the boundary/initial conditions via a number of collocation points 
(Eqs. (\ref{eq_loss_bc})(\ref{eq_loss_ic})), 
hard constraints that exactly enforce the boundary/initial conditions are also possible.
The idea is to design a problem specific ansatz that automatically satisfies 
the conditions of interest. More details about hard constraints in PINNs can be found in 
\cite{lagaris1998artificial, hao2024structure, mcfall2009artificial, lu2021physics, sukumar2022exact}.

\begin{figure}[H]
        \includegraphics[scale=0.55]{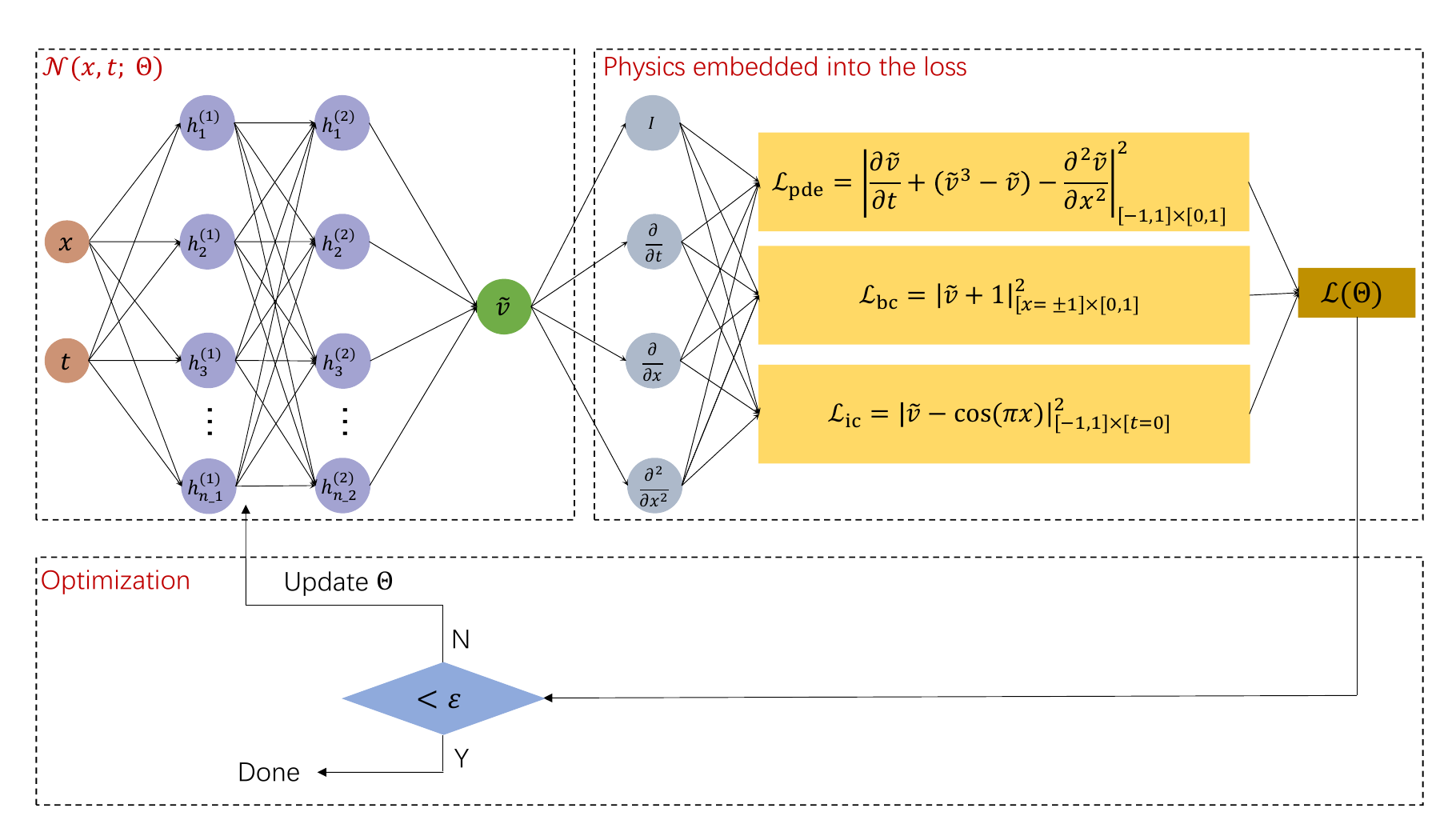}       
        \caption{An example of the baseline PINN. Here the PDE problem comprises a 
        one-dimensional (1D) Allen-Cahn equation and the boundary/initial conditions:  
        $\frac{\partial{v}}{\partial{t}} = -(v^3 - v) + \frac{\partial^2{v}}{\partial{x}^2}, 
        x \in [-1, 1], t \in [0, 1], v(x= \pm 1,t) = -1, v(x, t = 0) = \cos(\pi{x}).$ 
        In this figure, we use the simplified notations of the loss terms 
        as defined in Eq. (\ref{eq_loss_components}) for brevity.}
        \label{fig_simple_PINN}
 \end{figure}

With the loss function Eq. (\ref{eq_total_loss}), the network parameters $\Theta$ 
are determined by solving the optimization problem 
\begin{align}
        \Theta^{*} = \underset{\Theta}{\mathrm{argmin}} \quad {\mathcal{L}({\Theta})}.  
        \label{eq_optimization}
\end{align}
This can be done via gradient descent algorithms \cite{ruder2016overview}. 
Then the predicted solution to Eq. (\ref{eq_PDE_problem}) is 
$\bm{\tilde{v}}(\bm{x},t) = \mathcal{N}(\bm{x},t; \Theta^{*})$. 
It is pertinent to mention that to have a good solution, 
all terms in $\mathcal{L}(\Theta)$ should be small enough, 
so Eq. (\ref{eq_optimization}) is a multi-objective optimization problem  
and is probably difficult to solve \cite{rohrhofer2023apparent}.
The above explanations of PINNs may be visualized in Fig. \ref{fig_simple_PINN}, 
where an Allen-Cahn equation with prescribed boundary and initial 
conditions is used as an example of PDE problems.

PINNs are flexible, requiring a great effort
to choose suitable hyper-parameters, often by trial and error. 
However, the programming is normally not difficult, and can be 
implemented with the popular deep learning frameworks 
\href{https://www.tensorflow.org/}{TensorFlow} or 
\href{https://pytorch.org/}{PyTorch},  
or with particularly developed libraries, e.g.,
\href{https://developer.nvidia.com/modulus}{NVIDIA Modulus}, 
\href{https://docs.sciml.ai/NeuralPDE/stable/}{NeuralPDE.jl} \cite{Zubov2021NeuralPDE},
and \href{https://deepxde.readthedocs.io/en/latest/}{DeepXDE} \cite{lu2021deepxde}.

\section{Phase-field model of ferroelectric microstructure evolution}
\label{sect_phase_field_model}
Ferroelectric materials show spontaneous polarizations below the Curie temperature, 
and the polarizations can be changed by applying an electric or mechanical load. 
If over a region in the material, the polarizations orient in the same direction, 
this region is called a domain. Domains are the primary microstructure in ferroelectrics, 
influenced by both properties of the material and external loads. 
To simulate the evolution of ferroelectric domains, the phase-field method, 
a general framework to model spatiotemporal change of microstructural patterns 
in a wide range of materials processes based on the diffuse-interface approach 
\cite{moelans2008introduction} and thermodynamics \cite{chen2022classical} is used.
In this framework, polarizations are taken as the phase-field variable 
(i.e., the order parameter), and their evolution is governed by the 
TDGL equation. Besides, ferroelectric microstructure 
evolution is an electromechanical behavior, where the polarization field, 
strain field, and electric field interact with each other, so 
the equation of momentum balance and the Maxwell's equation to 
govern the latter two fields, respectively, are required as well. 
Finally, providing proper boundary and initial conditions, the mathematical model 
of ferroelectric microstructure evolution is constructed.  
We will present this model in a Cartesian coordinate system $O-x_1x_2x_3$, with 
$\bm{e}_1, \bm{e}_2, \bm{e}_3$ as the unit vectors in 1, 2, and 3 directions, respectively.

\subsection{Governing equations}
To model ferroelectric microstructure evolution using the phase-field method, 
the polarization $\bm{P} = P_i{\bm{e}_i}, i \in \{1,2,3\}$ 
is chosen as the order parameter. The Einstein summation notation is used here and thereafter. 
This order parameter is a non-conserved 
variable, and its evolution is described by the TDGL equation  
\begin{align}
       \frac{\partial{P_i}}{\partial{t}} 
    = -{\zeta}(\frac{\partial{\psi}}{\partial{P_i}} 
      - \frac{\partial}{\partial{x_j}}(\frac{\partial{\psi}}{\partial{P_{i,j}}})), 
      \: i,j \in \{1,2,3\} 
      \:\:\:\: \text{in} \:\: \Omega , \label{eq_TDGL}
\end{align}
where $\zeta$ is the kinetic coefficient related to the domain mobility, 
$P_{i,j} = \frac{\partial{P}_i}{\partial{x}_j}$, i.e., a comma in a subscript 
stands for spatial differentiation, and $\psi$ is the total free energy density 
(see in \nameref{sect_appendx_A} for the expression). 
The equation of momentum balance and the Maxwell's equations are 
\begin{align}
        &\sigma_{ij,j} = 0, \: i,j \in \{1,2,3\}   
        \:\:\:\: \text{in} \:\: \Omega , \label{eq_momentum}  \\
        & D_{i,i} = 0, \: i \in \{1,2,3\} 
        \:\:\:\: \text{in} \:\: \Omega ,  \label{eq_Maxwell}
\end{align}
where $\sigma_{ij}$ is the stress component, 
and $D_i$ is the electric displacement component. 
$\sigma_{ij}$ and $D_i$ are related to the strain component  
$\varepsilon_{ij} = \frac{1}{2}({u_{i,j}} + {u_{j,i}})$
and the electric field component $E_{i} = -\phi_{,i}$, 
respectively, through the constitutive laws given in \nameref{sect_appendx_B}. 
The (mechanical) displacement $\bm{u} = u_i{\bm{e}_i}$
and the electric potential $\phi$ are the other two unknown variables 
besides the polarization $\bm{P}$ for the phase-field model of interest.

\subsection{Boundary and initial conditions}
To obtain a specific solution to the model Eqs. (\ref{eq_TDGL})-(\ref{eq_Maxwell}), 
boundary and initial conditions must be prescribed. 
The surface gradient flux of polarizations is usually assumed to be zero, i.e.,  
\begin{align}
        &\frac{\partial{\psi}}{\partial{P_{i,j}}}n_j = \bar{\pi}_i = 0, 
        \quad i, j \in \{1,2,3\} \quad \text{on} \quad  \partial{\Omega}, 
        \label{eq_bc_gradient}
\end{align}
where $n_j$ is the component of a unit vector normal to the boundary $\partial{\Omega}$.
The mechanical and electric boundary conditions may be expressed as 
\begin{align}
        & \sigma_{ji}n_j = \bar{\tau}_i, 
        \; i, j \in \{1,2,3\} \; \text{on} \; \Gamma_{\text{N}}^{\text{m}}, 
        \quad \quad u_i = \bar{u}_i, \; i \in \{1,2,3\} \; \text{on} \;  \Gamma_{\text{D}}^{\text{m}}; 
        \label{eq_bc_mechanical}\\
        & {D_j}{n_j} = \bar{q}, \; j \in \{1,2,3\} \; \text{on} \; \Gamma_{\text{N}}^{\text{e}},
        \quad \quad \phi = \bar{\phi} \; \text{on} \;  \Gamma_{\text{D}}^{\text{e}}. 
        \label{eq_bc_electric}      
\end{align}
In Eqs. (\ref{eq_bc_mechanical})(\ref{eq_bc_electric}), 
$\Gamma_{\text{N}}^{\text{m}} \cap \Gamma_{\text{D}}^{\text{m}} = 
\Gamma_{\text{N}}^{\text{e}} \cap \Gamma_{\text{D}}^{\text{e}} =
\emptyset$, 
$\Gamma_{\text{N}}^{\text{m}} \cup \Gamma_{\text{D}}^{\text{m}} = 
\Gamma_{\text{N}}^{\text{e}} \cup \Gamma_{\text{D}}^{\text{e}} =
\partial{\Omega}$, and the subscripts $(\cdot)_{\text{N}}, (\cdot)_{\text{D}}$ 
denote the Neumann boundary and the Dirichlet boundary, respectively. 
The values of $\bar{\tau}_i$, $\bar{u}_i$, $\bar{q}$, and $\bar{\phi}$ 
depend on external loads. For instance, if a ferroelectric material 
is free of traction and open-circuited on all its surfaces, 
then we will have $\bar{\tau}_i = 0, \bar{q} = 0$, 
and $\Gamma_{\text{N}}^{\text{m}} = \Gamma_{\text{N}}^{\text{e}} = \partial{\Omega}$.

To completely close the PDE system Eqs. (\ref{eq_TDGL})-(\ref{eq_bc_electric}), 
an initial condition for the polarization $\bm{P}$ should be provided as below: 
\begin{align}
        &P_i = P_{i}^{\text{init}}, 
        \quad i \in \{1,2,3\} \quad \text{at} \quad t = 0 \quad \text{in} \quad \Omega, \label{eq_init}
\end{align}
where $P_{i}^{\text{init}}$ is a known spatial distribution  
of the polarization component $P_i$. 
We note that there are no initial conditions for the 
displacement $\bm{u}$ and the electric potential $\phi$ 
since the governing equations of $\bm{u}$ and $\phi$ 
(i.e., Eqs. (\ref{eq_momentum})-(\ref{eq_Maxwell})) 
do not have any temporal terms. Nevertheless, $\bm{u}$ and $\phi$ are temporally 
dependent fields because they are coupled with the polarization $\bm{P}$, 
which evolves until the equilibrium.

\section{The en-PF PINN approach: energy embedded into the PINN of the phase-field model}
\label{sect_energy_dissipation}
The phase-field model of ferroelectric microstructure evolution 
Eqs. (\ref{eq_TDGL})-(\ref{eq_init}) is a second-order PDE system 
with three coupled unknown fields $\bm{P}, \bm{u}$, and $\phi$, 
all spatially and temporally dependent. This model is nonlinear,  
because Eq. (\ref{eq_TDGL}) is highly nonlinear since 
$\psi_{\text{Landau}}$ (see in \nameref{sect_appendx_A})
is a sextic polynomial of the polarization. To solve such a 
time-dependent, nonlinear, and high-order PDE system of multi-physics 
with the PINN framework illustrated in Sect. \ref{sect_overview_of_pinn}  
is a tough task, and readers could refer to our previous work 
\cite{shang2024quantification} for more details. 

In this work, considering that the acquisition of steady microstructures is 
one central focus in simulations of ferroelectric microstructure evolution, 
we go directly to a static problem rather than fight against 
the dynamic problem Eqs. (\ref{eq_TDGL})-(\ref{eq_init}). 
This static problem is almost the same as the dynamic problem, 
but the left-hand side of Eq. (\ref{eq_TDGL}), namely 
$\frac{\partial{P_i}}{\partial{t}}$ is set to be zero, so the following equation 
\begin{align}
        0 = -{\zeta}(\frac{\partial{\psi}}{\partial{P_i}} 
       - \frac{\partial}{\partial{x_j}}(\frac{\partial{\psi}}{\partial{P_{i,j}}})), 
       \: i,j \in \{1,2,3\} 
       \:\:\:\: \text{in} \:\: \Omega , \label{eq_TIGL}
 \end{align}
is solved together with other governing PDEs (\ref{eq_momentum})-(\ref{eq_Maxwell}) 
and all boundary conditions (\ref{eq_bc_gradient})-(\ref{eq_bc_electric}). 
The initial condition Eq. (\ref{eq_init}) should not be involved for 
a static problem. This reformulation from the dynamic to the static is physically valid because 
\begin{itemize}
        \item {ferroelectric microstructure evolution of interest will eventually 
        reach the equilibrium where the microstructure stops to change in time, and} 
        \item {the steady microstructure is insensitive 
        \footnote{If applying different initial conditions $\mathsf{I_1}, \mathsf{I_2}$ 
        to a ferroelectric system, the system may show different steady microstructures 
        $\mathsf{M_1}, \mathsf{M_2}$, but the difference only consists in that 
        the polarizations in $\mathsf{M_1}$ have opposite directions to the 
        polarizations in $\mathsf{M_2}$. Here we ignore this difference, and in this sense, 
        the steady microstructure is insensitive to the initial condition. } 
        to the initial condition 
        Eq. (\ref{eq_init}).}
\end{itemize}
However, the static problem Eqs. (\ref{eq_TIGL})(\ref{eq_momentum})-(\ref{eq_bc_electric}) 
is ill-posed, i.e., likely to have many solutions. Once again take 
a ferroelectric material free of traction and open-circuited on all its surfaces 
as an example. In such a case, a trivial solution $\bm{P} = \bm{0}$, 
$\bm{u} = \bm{0}$, and $\phi = 0$ exists. Thus, to obtain a meaningful solution 
to the static problem, additional constraints beyond 
Eqs. (\ref{eq_TIGL})(\ref{eq_momentum})-(\ref{eq_bc_electric}) must be enforced. 

This is where the law of energy dissipation for ferroelectric microstructure 
evolution works its magic. The total free energy of the system is the volume 
integral of the free energy density $\psi$, denoted by 
$\Psi = \int_{\Omega} \psi \text{d}\bm{x}$. 
The variation of $\Psi$ with respect to (w.r.t.) time is \cite{su2015phase}
\begin{align}
        \frac{\text{d}\Psi}{\text{d}t} = \int_{\Omega}\frac{\delta{\Psi}}{\delta{P_i}}
        \frac{\partial{P_i}}{\partial{t}}\text{d}\bm{x} = \int_{\Omega}
        (\frac{\partial{\psi}}{\partial{P_i}} - \frac{\partial}{\partial{x_j}}
        (\frac{\partial{\psi}}{\partial{P_{i,j}}}))\frac{\partial{P_i}}{\partial{t}}\text{d}\bm{x} 
        = -\frac{1}{\zeta}\int_{\Omega}(\frac{\partial{P_i}}{\partial{t}})^2\text{d}\bm{x} 
        \leq 0,
        \label{eq_energy_dissipation}
\end{align}
where the functional derivative of $\Psi$ \cite{li2023phase} and 
the TDGL equation (\ref{eq_TDGL}) are used. Eq. (\ref{eq_energy_dissipation}) 
means that the total free energy $\Psi$ is non-increasing in time. 
In other words, the energy is constantly dissipated during the evolution process.
Therefore, the system reaches the attainable minimum energy state when 
the evolution has stopped. This can serve as the additional constraints  
we seek to solve the aforementioned static problem 
Eqs. (\ref{eq_TIGL})(\ref{eq_momentum})-(\ref{eq_bc_electric}).

The combination of Eqs. (\ref{eq_TIGL})(\ref{eq_momentum})-(\ref{eq_bc_electric}) 
with the constraint of attainable minimum energy creates a well-posed problem 
\footnote{We note that if two solutions contain vectors which have the same magnitude 
but opposite directions, e.g., $\bm{P} = (1,0,0)$ in the first solution, 
while $\bm{P} = (-1, 0, 0)$ in the second solution, and there are no other types of 
difference between the two solutions, the two solutions are still 
regarded as one unique solution. Under this assumption, the problem 
described here is well-posed.}, which could be difficult to tackle with FEM, 
but feasible with PINNs, thanks to the versatile formulations 
of PINN loss functions. The term associated with the total free energy 
is embedded into the loss function as   
\begin{align}
        \mathcal{L}_{\text{energy}} = {\exp}({\eta}(\Psi - \Psi^0)), 
        \label{eq_loss_total_energy}
\end{align}
where $\Psi^0$ represents the total free energy of a known state 
(e.g., the prescribed initial state), 
and $\eta$ is a positive constant. The total free energy $\Psi$ 
is a function w.r.t. the network parameters $\Theta$ because it is 
evaluated based on the unknown fields $\bm{P}$, $\bm{u}$, and $\phi$, 
which are approximated by the outputs of the network in a PINN framework.
Accounting for energy dissipation, 
at the final equilibrium state, we have $\Psi - \Psi^0 \leq 0$,   
thus $0 < \exp(\Psi - \Psi^0) \leq 1$, and the smaller $\Psi$ is, 
the closer $\exp(\Psi - \Psi^0)$ is to zero. 
In the implementation, the integral in Eq. (\ref{eq_loss_total_energy})
can be computed using the mean rule, namely 
\begin{align}
        \Psi = \int_{\Omega}\psi{\text{d}}{\bm{x}} \approx V \frac{1}{N_{\text{B}}}\sum_{i=1}^{N_{\text{B}}}\psi^{[i]}, 
\end{align}
where $N_{\text{B}}$ is the total number of collocation points in the domain $\Omega$, 
$V$ is volume of $\Omega$, and $\psi^{[i]}$ is the free energy density evaluated at the 
collocation point $\bm{x}^{[i]} = (x_1^{[i]}, x_2^{[i]}, x_3^{[i]})$. Other rules such as the trapezoidal rule and Simpson's rule 
are alternative choices to compute the integral \cite{nguyen2020deep}.

\begin{figure}[ht]
        \includegraphics[scale=0.55]{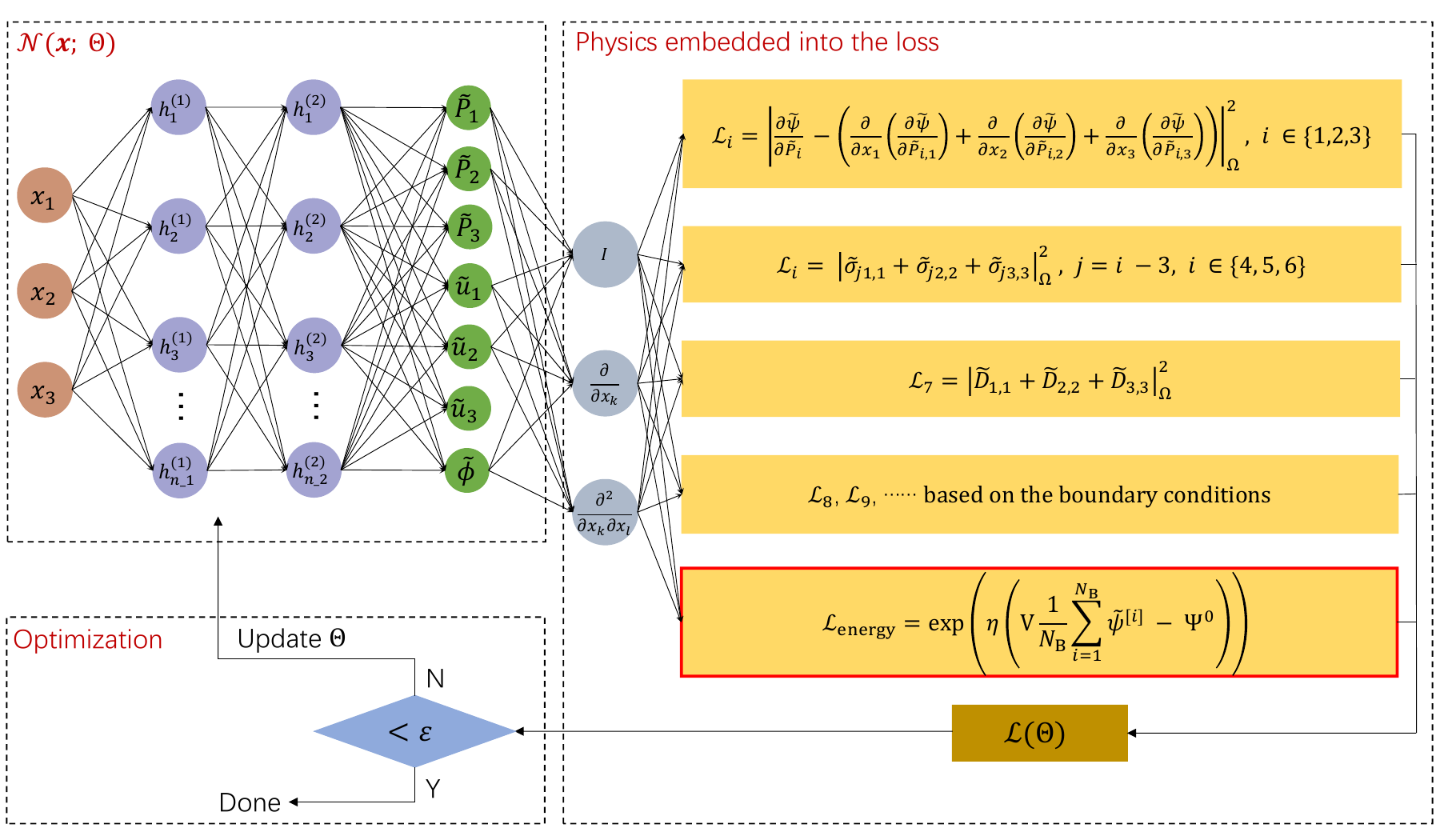}       
        \caption{The proposed PINN algorithm to obtain the steady ferroelectric microstructure.
        Firstly the dynamic problem of ferroelectric microstructure evolution is 
        reformulated to be a static problem, so the network inputs do not contain 
        time $t$, and no temporal terms present in $\mathcal{L}_i, i \in \{1, 2, 3\}$. 
        Then an energy term (i.e., $\mathcal{L}_{\text{energy}}$, 
        denoted in red) based on energy dissipation is added to the 
        loss function to avoid a trivial solution. 
        Finally, an optimization procedure is performed to determine 
        the network parameters $\Theta$. In this figure, $\tilde{\cdot}$ means that the  
        associated quantities are computed using the PINN predicted solution 
        $\tilde{P}_1, \tilde{P}_2, \tilde{P}_3,
        \tilde{u}_1, \tilde{u}_2, \tilde{u}_3,  \tilde{\phi}$.
        In this figure, we use the simplified notations of the loss terms 
        as defined in Eq. (\ref{eq_loss_components}) for brevity.}
        \label{fig_overview_3D}
 \end{figure}

In summary, we at first reformulate the phase-field model of ferroelectric microstructure 
evolution to be a static problem. This reformulation is physically valid, 
but the resulting static problem is not well-posed in mathematics. To obtain a meaningful 
solution for this static problem, we introduce an extra constraint inspired by the  
law of energy dissipation. This constraint is represented as an energy term 
indicated by Eq. (\ref{eq_loss_total_energy}) in the PINN loss function. 
We call this strategy as the en-PF PINN approach, a schematic illustration of which   
is shown in Fig. \ref{fig_overview_3D}. We note that in the en-PF PINN approach, 
the energy term Eq. (\ref{eq_loss_total_energy}) is a complement 
rather than a substitution to the governing equations. 
In other words, a standalone Eq. (\ref{eq_loss_total_energy}) 
combined with all necessary boundary conditions is not sufficient to determine 
a meaningful solution. That is why the loss function in Fig. \ref{fig_overview_3D} 
must contain all the terms derived from Eqs. (\ref{eq_TIGL})(\ref{eq_momentum})(\ref{eq_Maxwell}) 
(i.e., $\mathcal{L}_i, i \in \{1,2, \cdots, 7\}$ in Fig. \ref{fig_overview_3D}).
 
\section{Validation of the en-PF PINN approach}
\label{sect_examples}
This section devotes to validating the en-PF PINN approach through several examples. 
The PINN solution is compared with the reference solution 
obtained from FEM. In the FEM simulations, the dynamic problem  
Eqs. (\ref{eq_TDGL})-(\ref{eq_init}) is solved, where the 
initial values for each polarization component $P_i$ are  
sampled from a random normal distribution 
\footnote{The mean of the random normal distribution is set to be 0, 
and the standard deviation is set to be 0.001. Choosing the initial 
values for polarizations from a random distribution is the 
common practice, e.g., papers \cite{schrade2007domain, wang2009three, liu2019isogeometric}, 
in FEM simulations of ferroelectric microstructure evolution.}. 
The relative error used to evaluate the accuracy of 
the PINN solution $\tilde{v}$ against the FEM solution $\hat{v}$ is 
\begin{align}
        \mathcal{E}(\tilde{v}) = \frac{\sqrt{\sum(\tilde{v}^{[i]} - \hat{v}^{[i]})^2}}{\sqrt{\sum({\hat{v}^{[i]}})^2}}.
        \label{eq_relative_error} 
\end{align}

The weak form for the FEM simulations is given in \nameref{sect_appendx_B}.
Material parameters chosen to mimic ferroelectric microstructure evolution 
are provided in \nameref{sect_appendx_C}.
The PINN code is available from the github repository 
\href{https://github.com/whshangl/Open-en-PF-PINN}{here},
written using the open-source library 
\href{https://deepxde.readthedocs.io/en/latest/}{DeepXDE} 
within the backend \href{https://pytorch.org/}{PyTorch}. 
In this work, we run the PINN code on a computer 
that has one NVIDIA GeForce RTX 4090 GPU.

\subsection{Two examples in 2D}
\label{sect_2D_example}
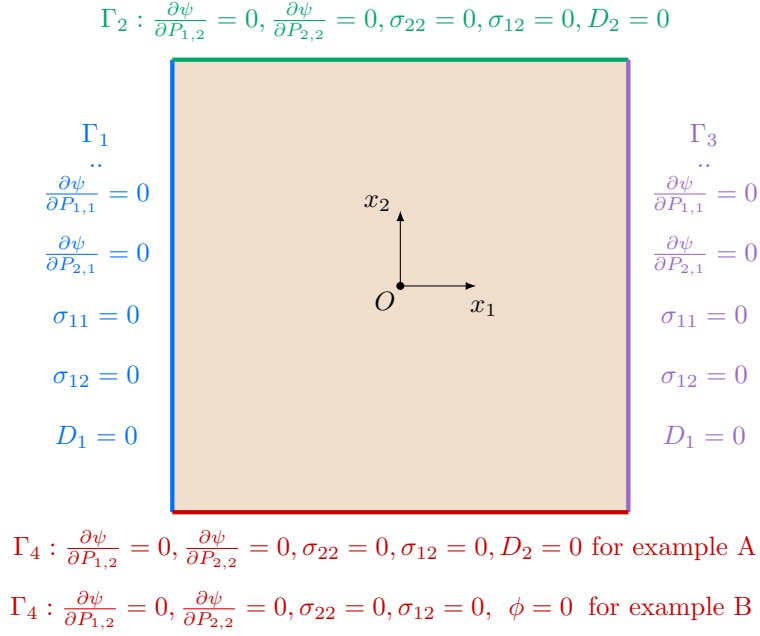
\begin{figure}[ht]
        \centering   
        \tikzset{every picture/.style={line width=1pt}}
        \begin{tikzpicture}

                \fill[almond] (0,0) rectangle (6,6);
                \draw[ultra thick, brandeisblue] (0,0)--(0,6);
                \draw[ultra thick, jade] (0,6)--(6,6);
                \draw[ultra thick, amethyst] (6,6)--(6,0);
                \draw[ultra thick, bostonuniversityred] (6,0)--(0,0);

                %%% coordinate system
                \fill[black](3,3)circle(0.15em);
                \draw[black, thin, -latex](3,3)--(4,3);
                \draw[black, thin, -latex](3,3)--(3,4);
                \draw[black](2.8,2.8) node {$O$};
                \draw[black](4.1,2.7) node {$x_1$};
                \draw[black](2.7,4.1) node {$x_2$};

                \draw[thick, brandeisblue](-1, 5) node {$\Gamma_1$}; 
                \draw[thick, brandeisblue](-1, 4.6) node {..};
                \draw[thick, brandeisblue](-1, 4.2) node {$\frac{\partial{\psi}}{\partial{P}_{1,1}} = 0$};
                \draw[thick, brandeisblue](-1, 3.4) node {$\frac{\partial{\psi}}{\partial{P}_{2,1}} = 0$};
                \draw[thick, brandeisblue](-1, 2.6) node {$\sigma_{11} = 0$};
                \draw[thick, brandeisblue](-1, 1.8) node {$\sigma_{12} = 0$};
                \draw[thick, brandeisblue](-1, 1.0) node {$D_1 = 0$};

                \draw[thick, jade](2.8, 6.5) node {$\Gamma_2: 
                \frac{\partial{\psi}}{\partial{P}_{1,2}} = 0,  
                \frac{\partial{\psi}}{\partial{P}_{2,2}} = 0,  
                \sigma_{22} = 0,  
                \sigma_{12} = 0, 
                D_2 = 0$};

                \draw[thick, amethyst](7, 5) node {$\Gamma_3$};
                \draw[thick, amethyst](7, 4.6) node {..};
                \draw[thick, amethyst](7, 4.2) node {$\frac{\partial{\psi}}{\partial{P}_{1,1}} = 0$};
                \draw[thick, amethyst](7, 3.4) node {$\frac{\partial{\psi}}{\partial{P}_{2,1}} = 0$};
                \draw[thick, amethyst](7, 2.6) node {$\sigma_{11} = 0$};
                \draw[thick, amethyst](7, 1.8) node {$\sigma_{12} = 0$};
                \draw[thick, amethyst](7, 1.0) node {$D_1 = 0$};

                \draw[thick, bostonuniversityred](2.8, -0.5) node {$\Gamma_4:
                \frac{\partial{\psi}}{\partial{P}_{1,2}} = 0,  
                \frac{\partial{\psi}}{\partial{P}_{2,2}} = 0,  
                \sigma_{22} = 0,  
                \sigma_{12} = 0, 
                D_2 = 0$ for example A};

                \draw[thick, bostonuniversityred](2.75, -1.3) node {$\Gamma_4:
                \frac{\partial{\psi}}{\partial{P}_{1,2}} = 0,  
                \frac{\partial{\psi}}{\partial{P}_{2,2}} = 0,  
                \sigma_{22} = 0,  
                \sigma_{12} = 0, \: \:
                \phi = 0 \: $ for example B};
        \end{tikzpicture}
\caption{Boundary conditions for the two examples in the 2D case. For the example A, 
the boundary conditions are  $\bar{\pi}_i = 0, \bar{\tau}_i = 0, \bar{q} = 0$ 
on $\partial{\Omega}$ ($\partial{\Omega} = \Gamma_1 \cup \Gamma_2 \cup \Gamma_3 \cup \Gamma_4$). 
For the example B, the boundary conditions are 
$\bar{\pi}_i = 0, \bar{\tau}_i = 0$ on $\partial{\Omega}$, 
$\bar{q} = 0$ on $\partial{\Omega}\setminus{\Gamma_{4}}$, 
$\bar{\phi} = 0$ on $\Gamma_4$. }
\label{fig_2d_bcs}  
\end{figure}

To validate the en-PF PINN approach at a low computational expense, 
we at first deal with the 2D problem. The 2D problem is reduced  
from the phase-field model presented in Section \ref{sect_phase_field_model} 
by assuming $\varepsilon_{13} = \varepsilon_{23} = \varepsilon_{33} = 0$, 
and $P_3 = 0, E_3 = 0$. The computational domain $\Omega$ is a 
4 nm $\times$ 4 nm square in the plane $O-{x_1}{x_2}$. The origin $O$ 
is located at the center of the square, thus $x_1 \in [-2, 2], \: x_2 \in  [-2, 2]$.
Two examples, A and B with different boundary conditions are considered, 
as seen in Fig. \ref{fig_2d_bcs}. The network for the 2D cases has 2 nodes 
in the input layer for $x_1, x_2$, and 5 nodes in the output layer for 
$\tilde{P}_1, \tilde{P}_2, \tilde{u}_1, \tilde{u}_2, \tilde{\phi}$. 

\subsubsection{Loss function }
For the two cases, the loss function derived from Eqs. (\ref{eq_TIGL})
(\ref{eq_momentum})-(\ref{eq_bc_electric}) in the form of Eqs. (\ref{eq_loss_pde})
(\ref{eq_loss_bc})(\ref{eq_loss_total_energy}) is 
\begin{align}
        \mathcal{L}_{\text{2D}}(\Theta) = \sum_{i = 1}^{15}{\lambda_i}{\mathcal{L}_i}
        + {\mathcal{L}_{\text{energy}}}, 
        \label{eq_loss_2D}
\end{align}
where using the simplified notations of the loss terms as defined in Eq. (\ref{eq_loss_components}), we have 
\begin{subequations}
        \begin{align}
                &\mathcal{L}_i =\bigg |\frac{\partial{\tilde{\psi}}}{\partial{\tilde{{P}}_i}} 
                                - \left(\frac{\partial}{\partial{x_1}}(\frac{\partial{\tilde{\psi}}}{\partial{\tilde{{P}}_{i,1}}})
                                + \frac{\partial}{\partial{x_2}}(\frac{\partial{\tilde{\psi}}}{\partial{\tilde{{P}}_{i,2}}})\right)\bigg |_{\Omega}^2, 
                                \quad i \in \{1, 2\}, 
                                \label{eq_loss_TIGL_2D}\\
               &\mathcal{L}_i = |\tilde{\sigma}_{j1,1} + \tilde{\sigma}_{j2,2} |_{\Omega}^2,   
                                \quad j = i - 2, \quad i \in \{3, 4\},  
                                \label{eq_loss_mechanic_2D} \\
               &\mathcal{L}_5 = |\tilde{D}_{1,1} + \tilde{D}_{2,2} |_{\Omega}^2, 
                                \label{eq_loss_Maxwell_2D} \\
               &\mathcal{L}_i =\bigg |\frac{\partial{\tilde{\psi}}}{\partial{\tilde{{P}}_{j,1}}}\bigg |_{\Gamma_{1} \cup \Gamma_3}^2, 
                                 \: j = i - 5, \: i \in \{6, 7\}, \quad 
                \mathcal{L}_i =\bigg |\frac{\partial{\tilde{\psi}}}{\partial{\tilde{{P}}_{j,2}}}\bigg |_{\Gamma_{2} \cup \Gamma_4}^2, 
                                 \: j = i - 7, \: i \in \{8, 9\},  
                                 \label{eq_loss_grad_2D}\\
               &\mathcal{L}_{10} = |\tilde{\sigma}_{11} |_{\Gamma_1 \cup \Gamma_3}^2, \quad 
                \mathcal{L}_{11} = |\tilde{\sigma}_{22} |_{\Gamma_2 \cup \Gamma_4}^2, \quad 
                \mathcal{L}_{12} = |\tilde{\sigma}_{12} |_{\partial{\Omega}}^2, \label{eq_loss_traction_2D}\\
               &\mathcal{L}_{13} = |\tilde{D}_1 |_{\Gamma_1 \cup \Gamma_3}^2, \quad
                \mathcal{L}_{14} = |\tilde{D}_2 |_{\Gamma_2}^2, \quad 
                \mathcal{L}_{15} = 
                \begin{cases}
                        |\tilde{D}_2 |_{\Gamma_4}^2  \quad  \text{for the example A} \\  
                        |\tilde{\phi} |_{\Gamma_4}^2  \quad \text{for the example B}
                \end{cases}. \label{eq_loss_electric_2D}
        \end{align}
        \label{eq_loss_terms_2D}
 \end{subequations}
For the example A, the loss terms $\mathcal{L}_{14}$ and $\mathcal{L}_{15}$ can be merged into one, 
so its loss function actually has 15 terms in total. 
When computing $\mathcal{L}_\text{energy}$ in Eq. (\ref{eq_loss_2D}) 
based on Eq. (\ref{eq_loss_total_energy}), the value of $\Psi^0$ 
is required at first. Apparently the state $\bm{P} = \bm{0}, \bm{u} = \bm{0}, \phi = 0$ 
is an admissible state in the course of ferroelectric microstructure evolution, 
so $\Psi^0 = 0$ is an appropriate choice.

\subsubsection{Results of the example A}
\label{sect_result_example_A}
\begin{figure}[H]
        \centering 
        { 
        \includegraphics[scale=0.31]{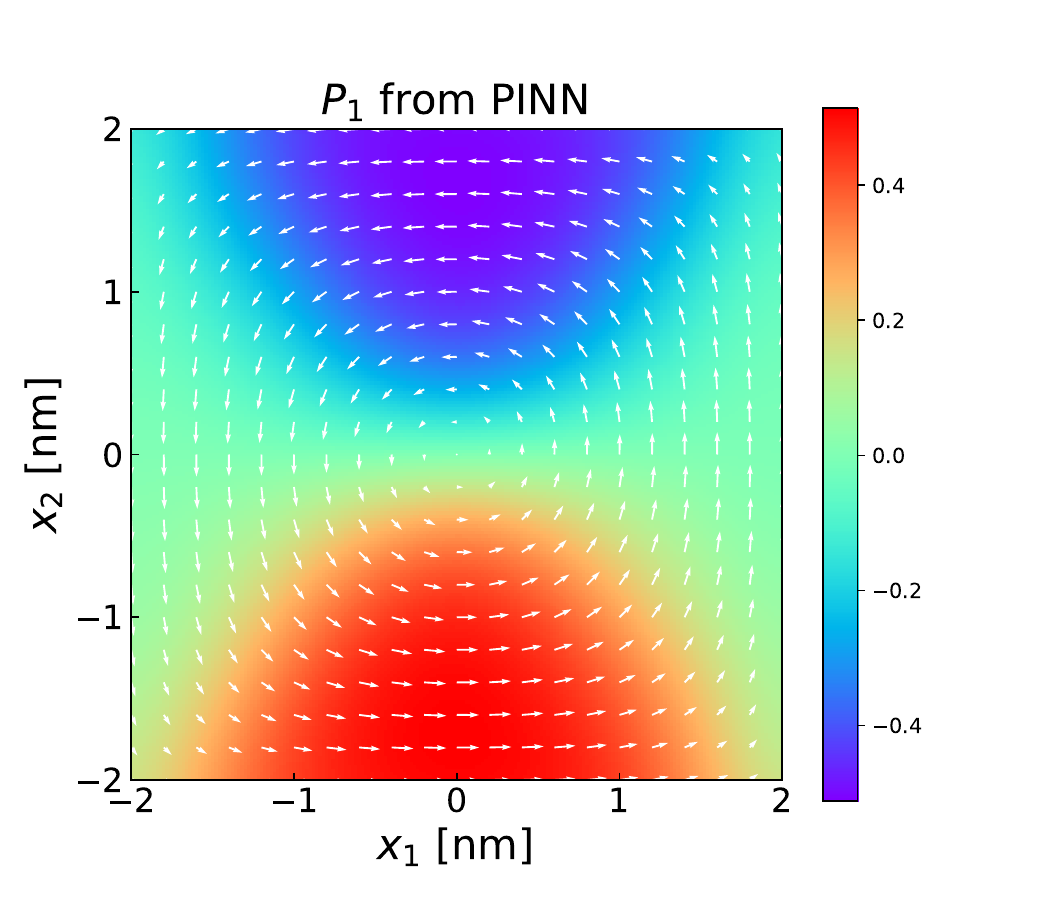} 
        \hspace{-0.6 cm} 
        \includegraphics[scale=0.31]{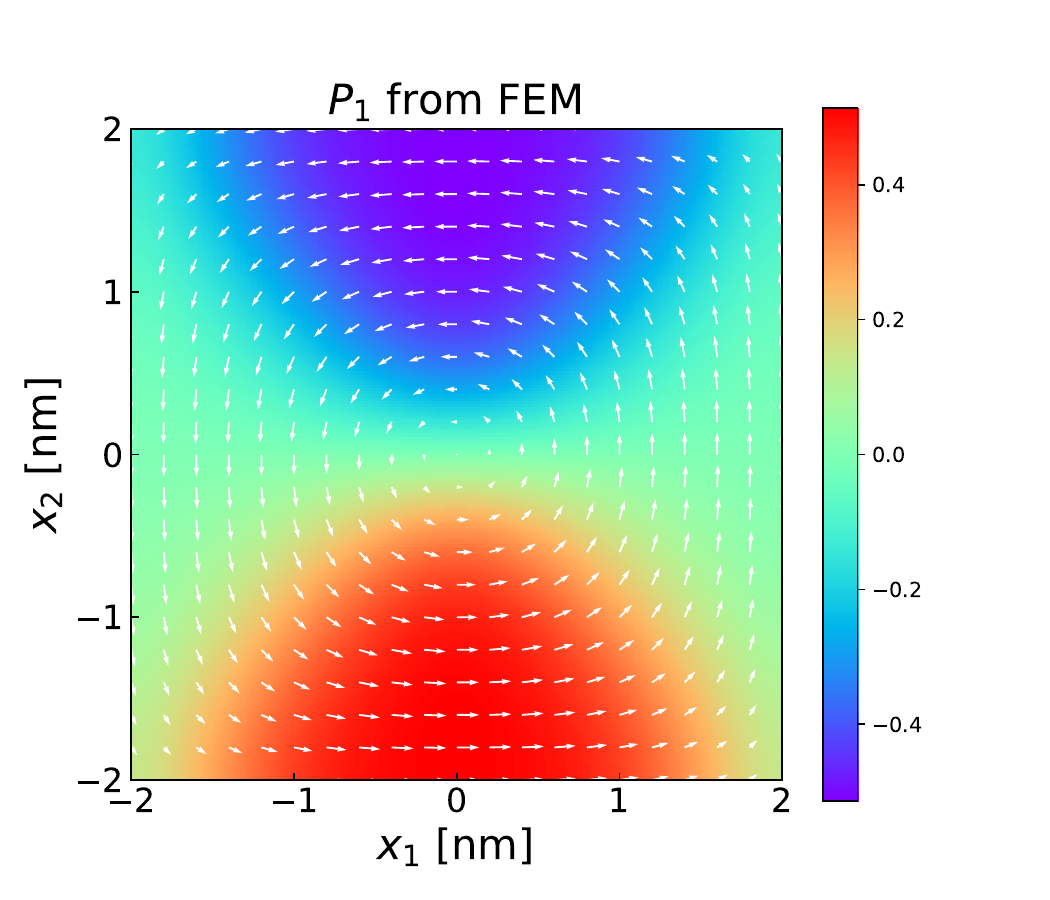}
        \hspace{-0.6 cm}
        \includegraphics[scale=0.31]{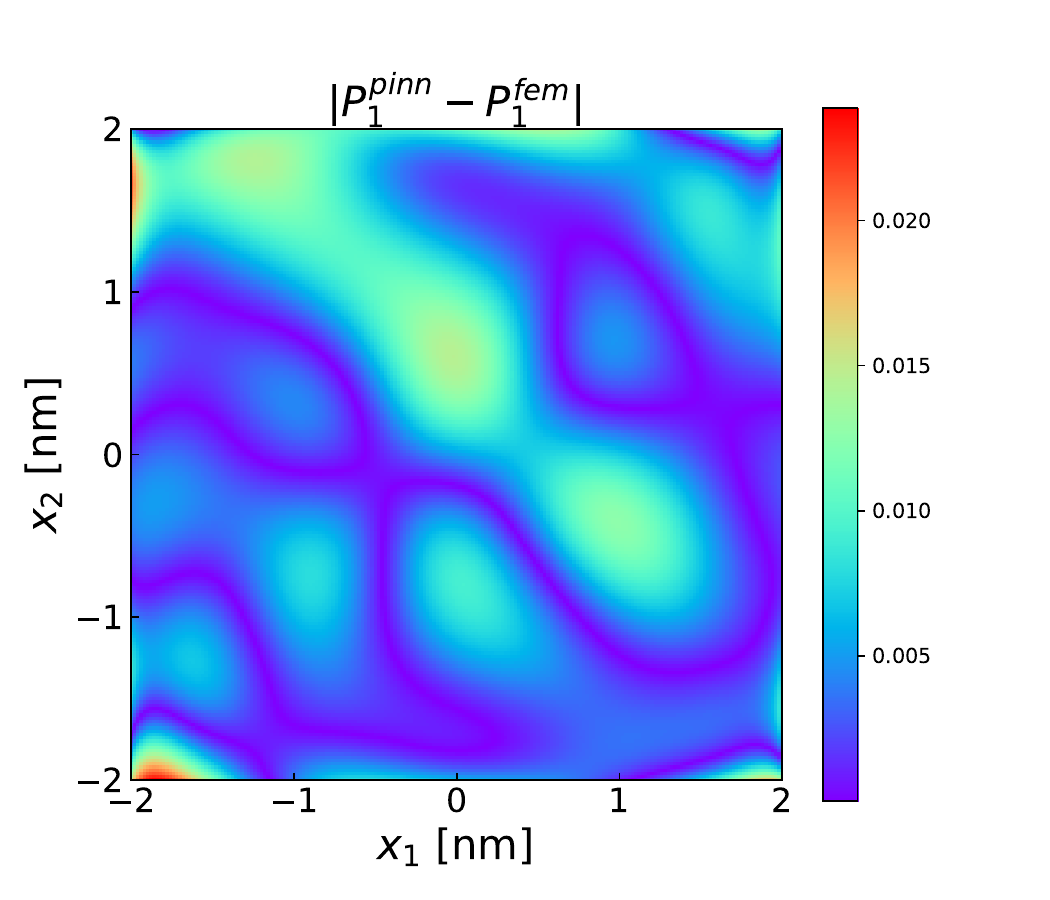}
        }
        \label{fig_P1_A}
        { 
        \includegraphics[scale=0.31]{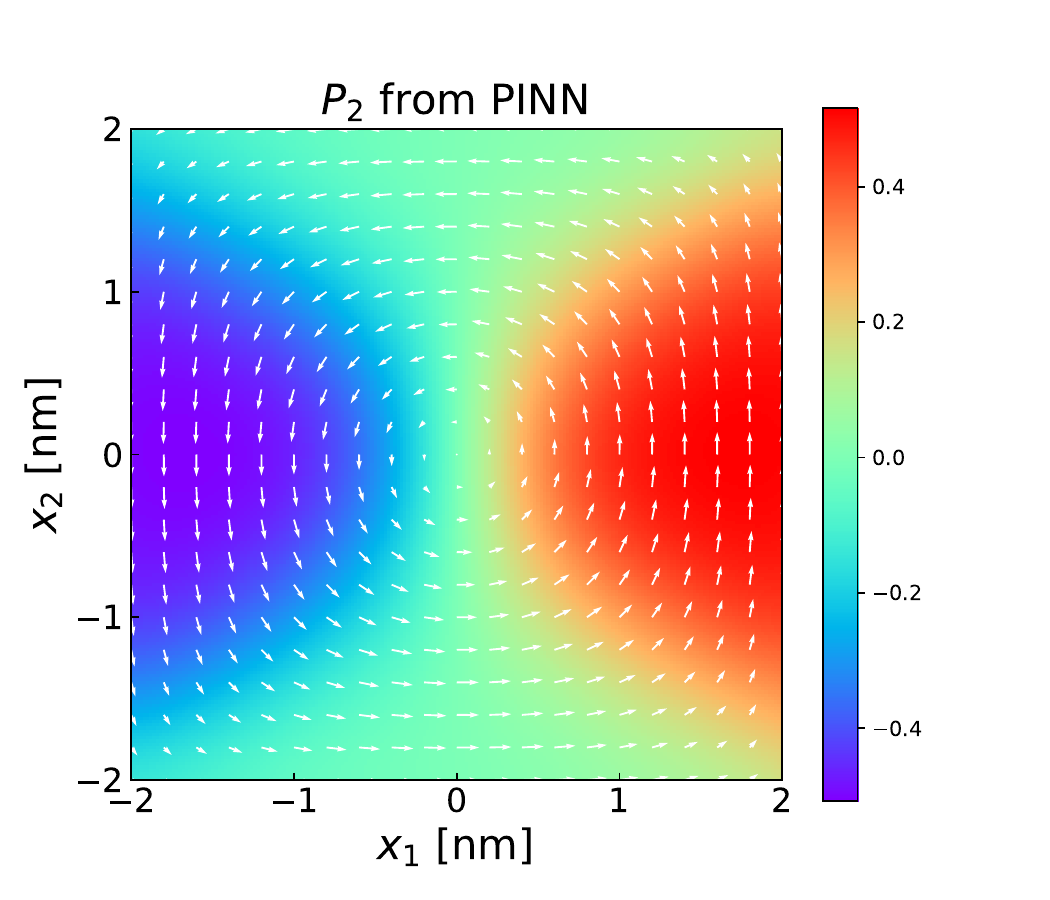} 
        \hspace{-0.6 cm}  
        \includegraphics[scale=0.31]{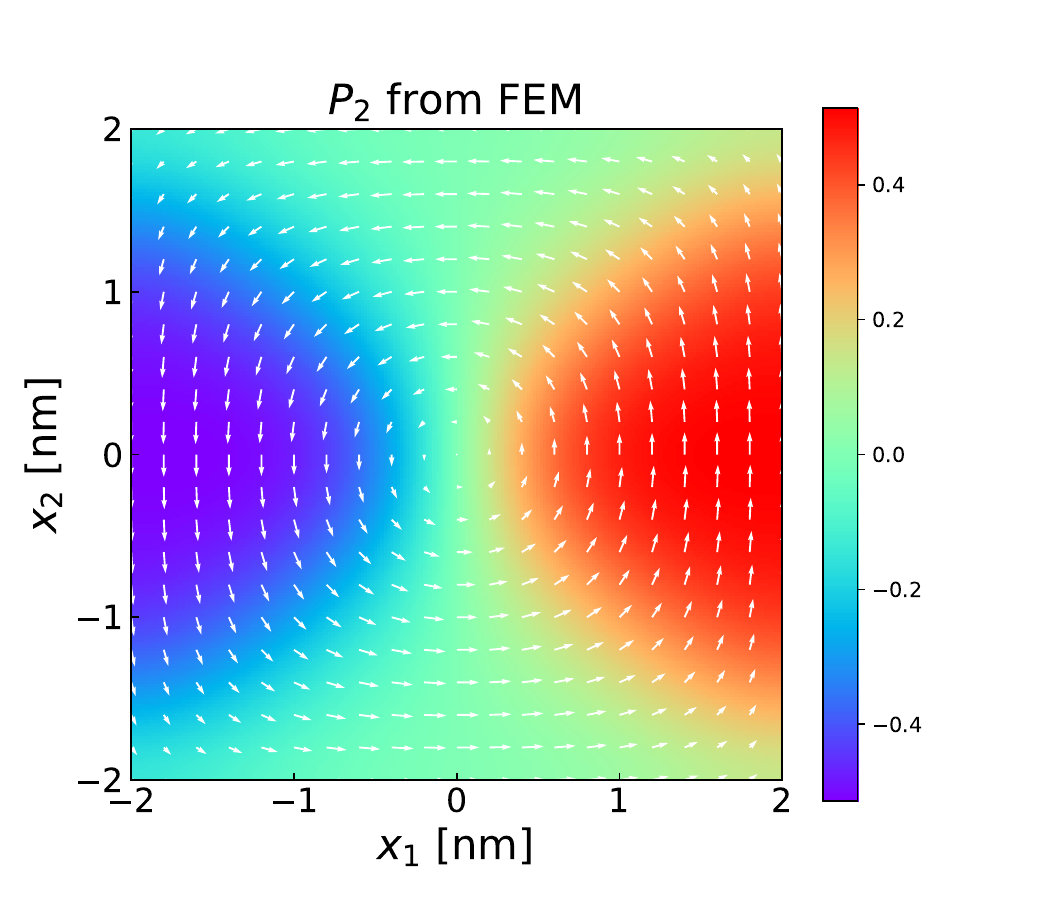}
        \hspace{-0.6 cm} 
        \includegraphics[scale=0.31]{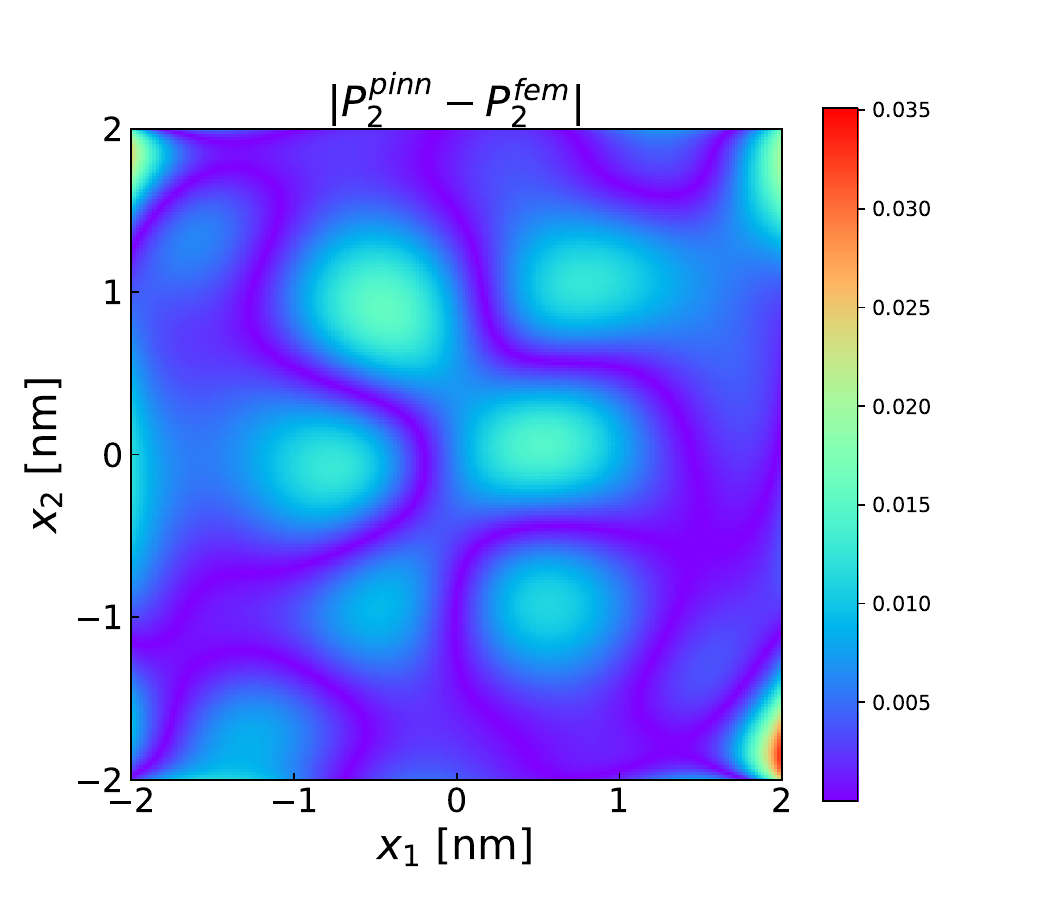}
        }
        \label{fig_P2_A}
        \vspace{-0.3 cm}
        \caption{Comparison of the polarization components 
        $P_1, \: P_2$ between the PINN and FEM results for the example A. 
        Arrows in the figures denote the polarization vector $\bm{P}$.} 
        \label{fig_P1P2_A}     
\end{figure}

A network with 3 hidden layers and 20 nodes in each hidden layer 
is trained to predict the steady ferroelectric microstructure of the example A. 
The weights $\lambda_i$ for the loss terms in Eq. (\ref{eq_loss_2D}) are chosen to be 
$\lambda_i = 1$ when $i \in \{1, 2, 3, 4\}$, $\lambda_5 = 100$, $\lambda_i = 10$ when  
$i \in \{ 6, 7, 8, \cdots, 15 \}$. The positive constant $\eta$ in the energy term 
$\mathcal{L}_\text{energy}$ is 1. The loss function Eq. (\ref{eq_loss_2D}) 
is evaluated using 11000 collocation points sampled uniformly from $\Omega$. 
To minimize the loss, 10000 epochs through the optimizer 
\href{https://pytorch.org/docs/stable/generated/torch.optim.Adam.html}{Adam} with the 
learning rate $10^{-3}$, followed by 10000 epochs through the optimizer 
\href{https://pytorch.org/docs/stable/generated/torch.optim.LBFGS.html}{LBFGS}, 
are implemented. The time spent on the training of such a PINN is about 8 minutes. 
The predicted results of $P_1, P_2$ are shown in Fig. \ref{fig_P1P2_A}, 
and the relative errors are $\mathcal{E}(\tilde{P}_1)$ = 1.96$\%$, $\mathcal{E}(\tilde{P}_2)$ = 2.02$\%$. 
Hence, it is clear that the PINN solution is in good agreement with the FEM solution.

In this example, $\eta = 1$ happens to be a good choice for the PINN, and 
other values for $\eta$, e.g., $\eta = 10$ will also work well,  
but it does not mean that $\eta$ is a redundant parameter. 
A detailed discussion of $\eta$ will be given in the example B.

\subsubsection{Results of the example B}
\begin{figure}[H]
        \centering 
        { 
        \includegraphics[scale=0.31]{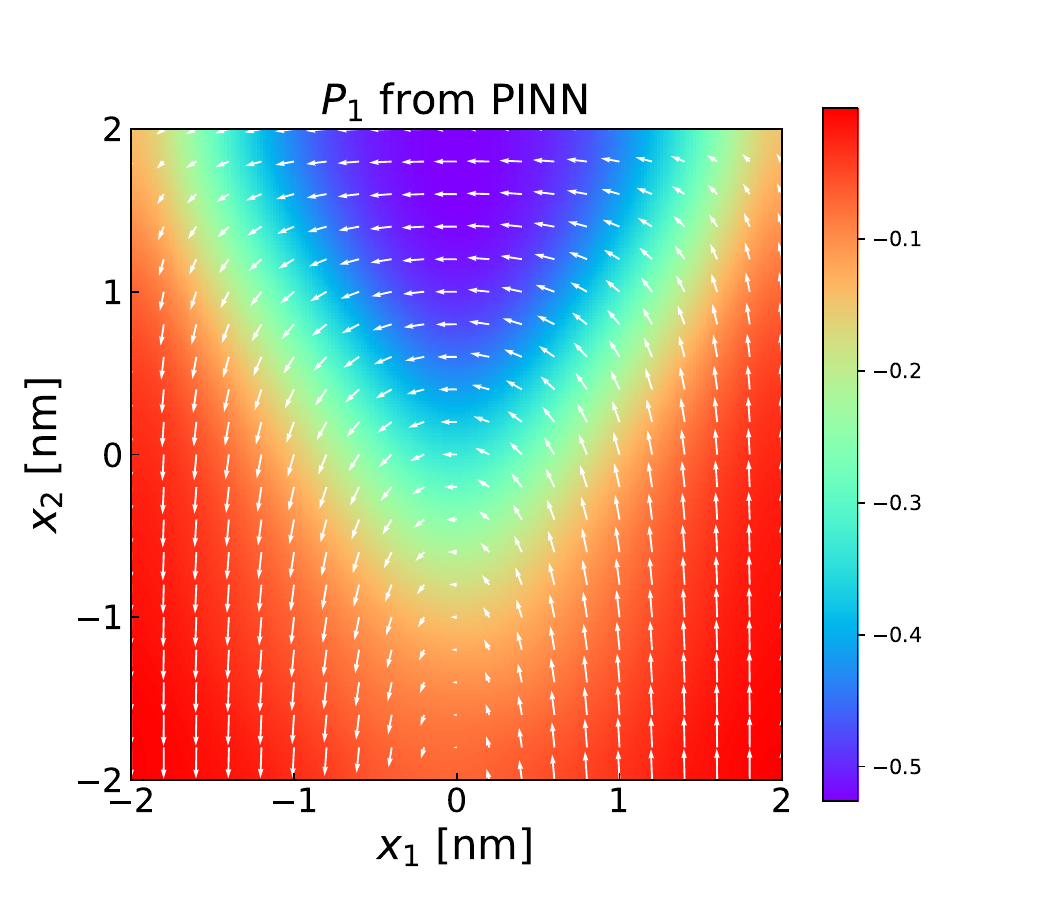} 
        \hspace{-0.6 cm} 
        \includegraphics[scale=0.31]{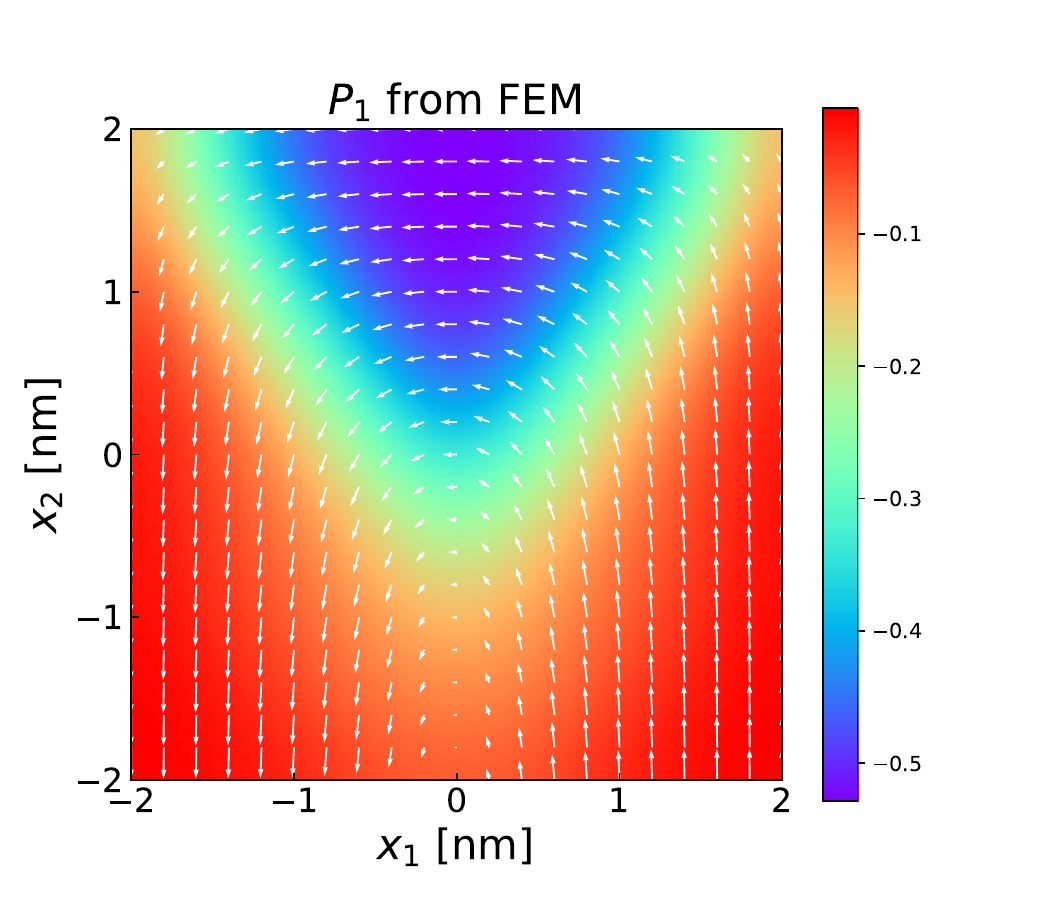}
        \hspace{-0.6 cm}
        \includegraphics[scale=0.31]{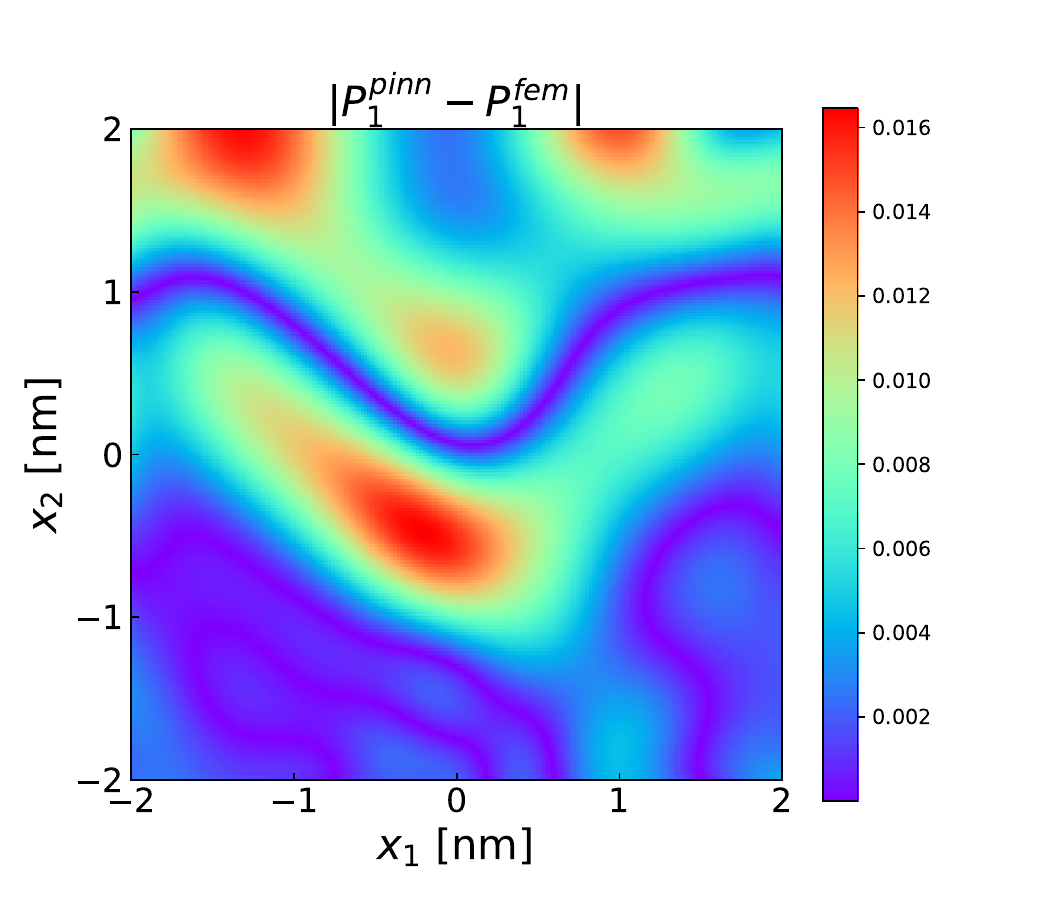}
        }
        \label{fig_P1_B}
        { 
        \includegraphics[scale=0.31]{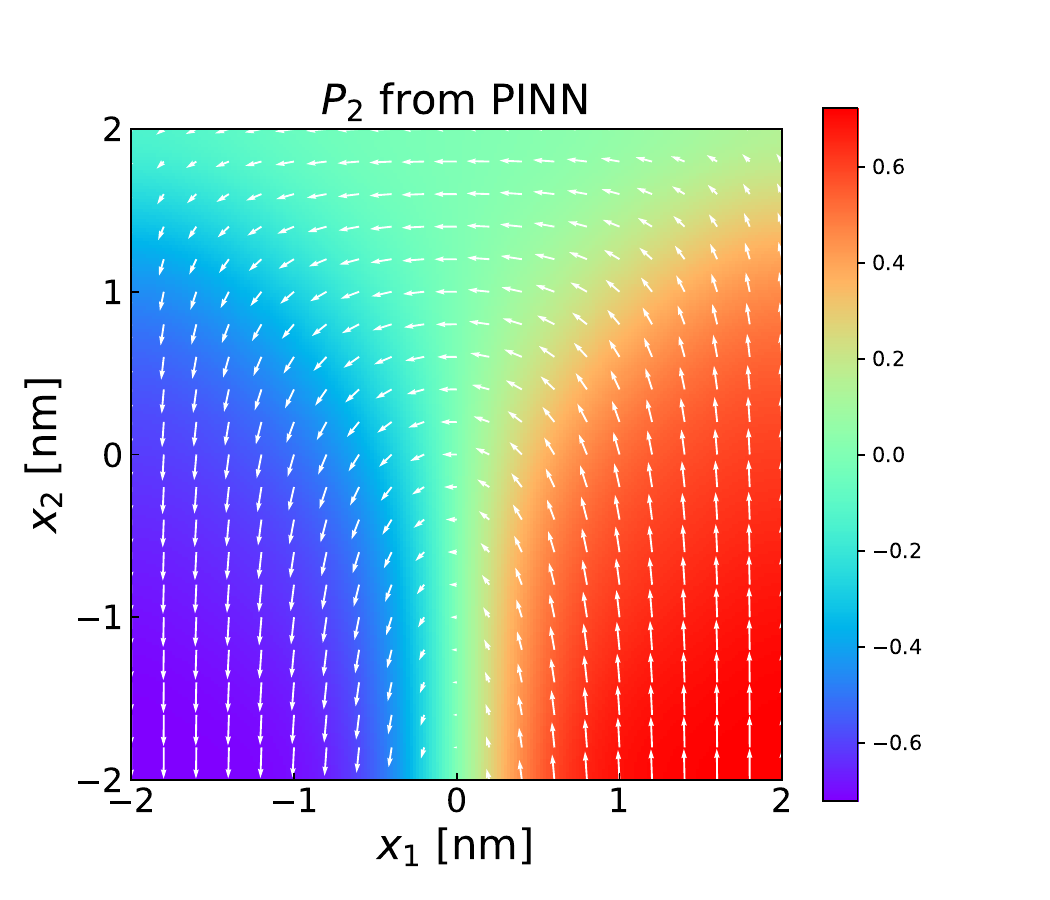} 
        \hspace{-0.6 cm}  
        \includegraphics[scale=0.31]{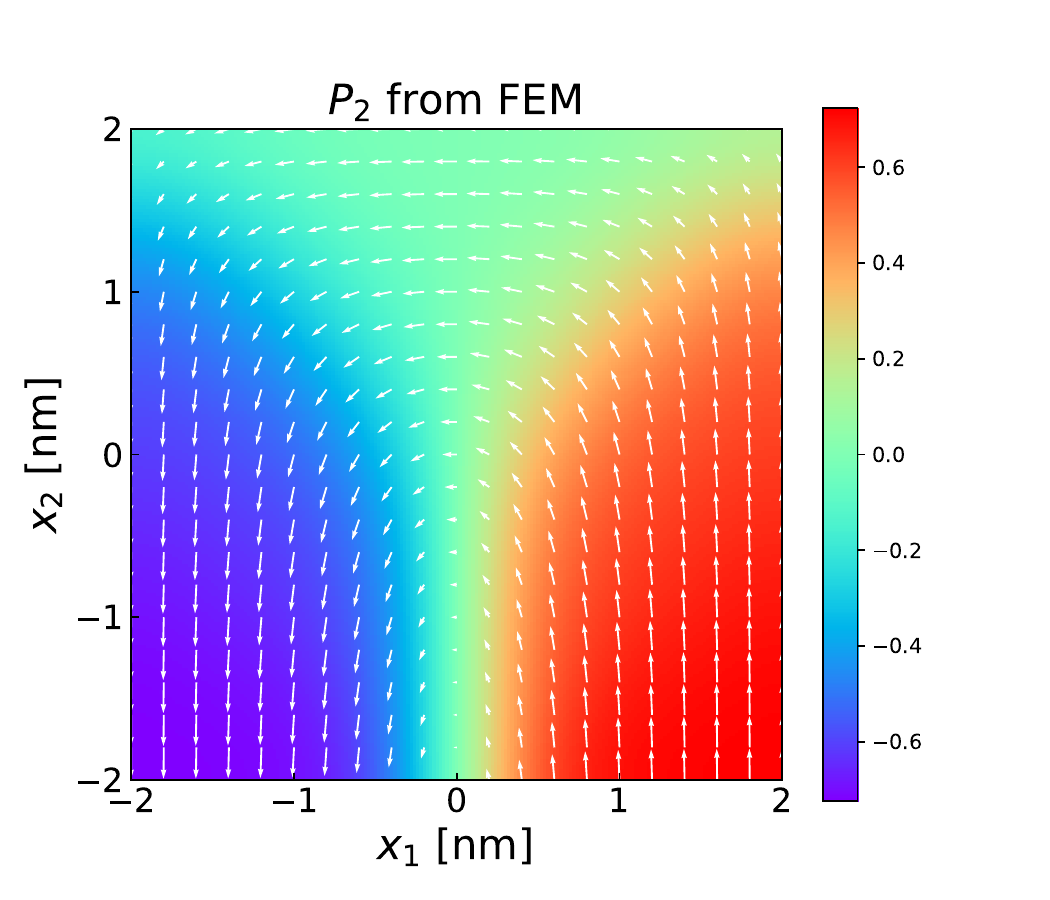}
        \hspace{-0.6 cm} 
        \includegraphics[scale=0.31]{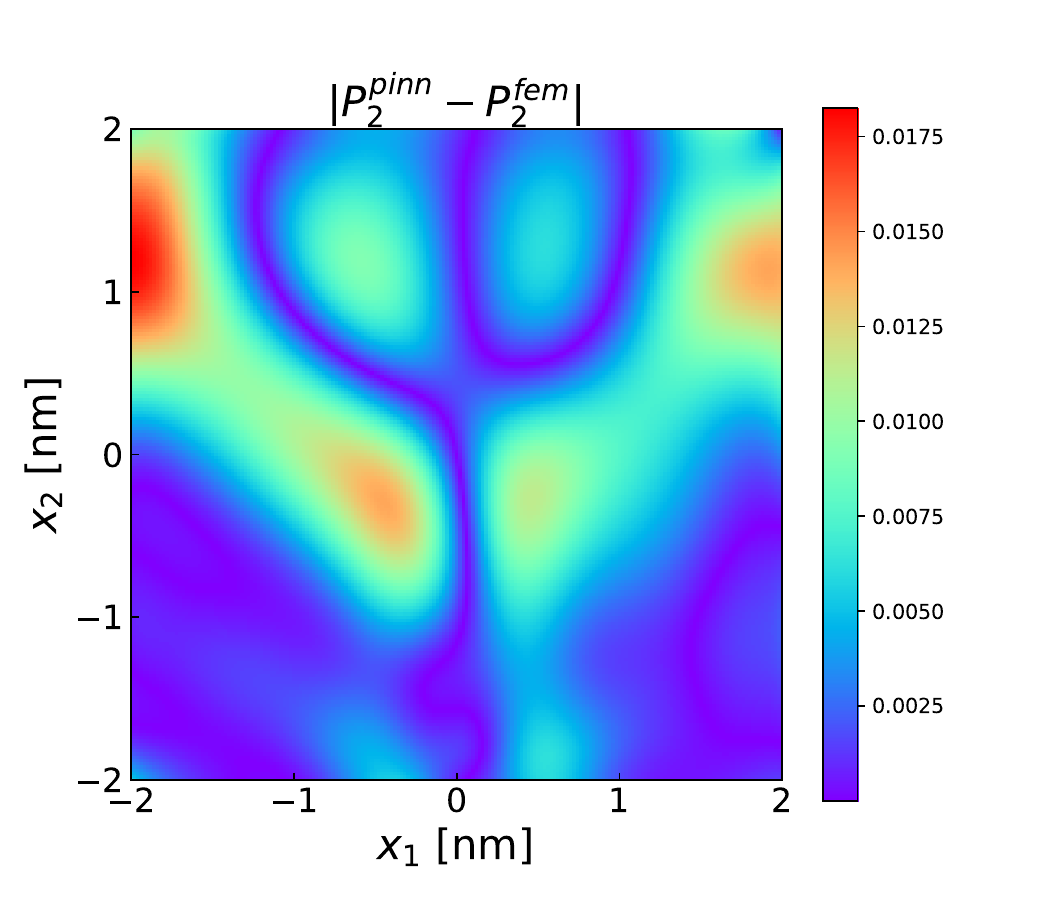}
        }
        \label{fig_P2_B}
        \vspace{-0.3 cm}
        \caption{Comparison of the polarization components $P_1, \: P_2$ 
        between the PINN and FEM results for the example B.   
        The direction of the polarization vector $\bm{P}$ obtained from the PINN 
        is opposite to that obtained from FEM, so here the 
        PINN figures are plotted using the opposite numbers of the raw PINN data.} 
        \label{fig_P1P2_B}     
\end{figure}

In this example, we use a network that has 3 hidden layers, 
each with 32 nodes to predict the the steady ferroelectric microstructure. 
The weights $\lambda_i$ for the loss terms in Eq. (\ref{eq_loss_2D}) are 
the same as those in the example A. The parameter $\eta$ is 20. 
The number of the collocation points is still 11000, while the 
training is a bit harder: the loss is minimized at first through the optimizer 
\href{https://pytorch.org/docs/stable/generated/torch.optim.Adam.html}{Adam} 
for 25000 epochs, where 5000, 5000, and 15000 epochs with the learning rates 
$10^{-2}$, $10^{-3}$, and $10^{-4}$, respectively, and then through the 
optimizer \href{https://pytorch.org/docs/stable/generated/torch.optim.LBFGS.html}{LBFGS} 
for 15000 epochs. It takes about 18 minutes to train this PINN. 
For this example, the PINN solution also agrees well with the FEM solution, 
as seen in Fig. \ref{fig_P1P2_B}. The relative errors 
are $\mathcal{E}(\tilde{P}_1)$ = 2.63$\%$, $\mathcal{E}(\tilde{P}_2)$ = 1.32$\%$.

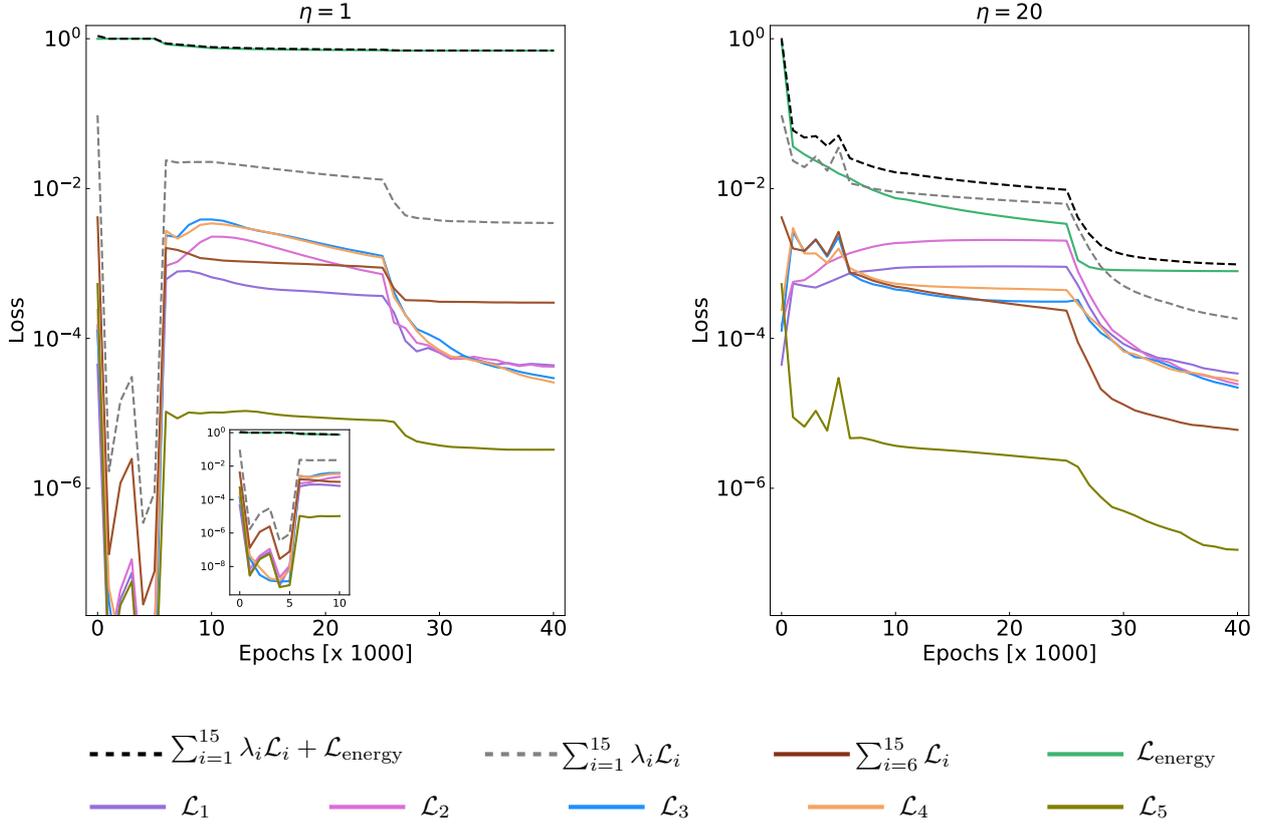
\begin{figure}[H]
\begin{tikzpicture}
        \node[inner sep=0pt] (russell) at (0,0)
                {\includegraphics[scale=0.4]{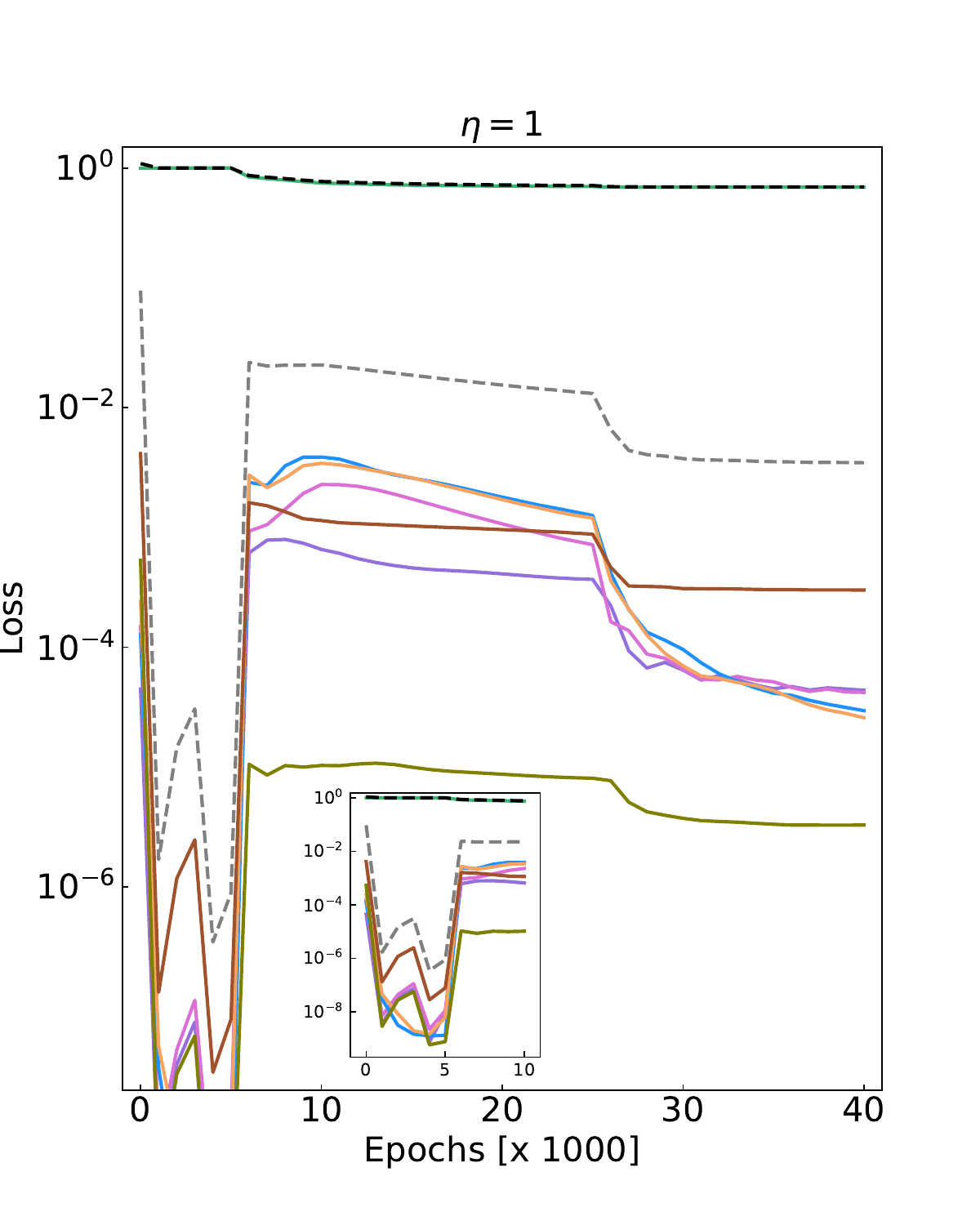}};
        \node[inner sep=0pt] (whitehead) at (9,0)
                {\includegraphics[scale=0.4]{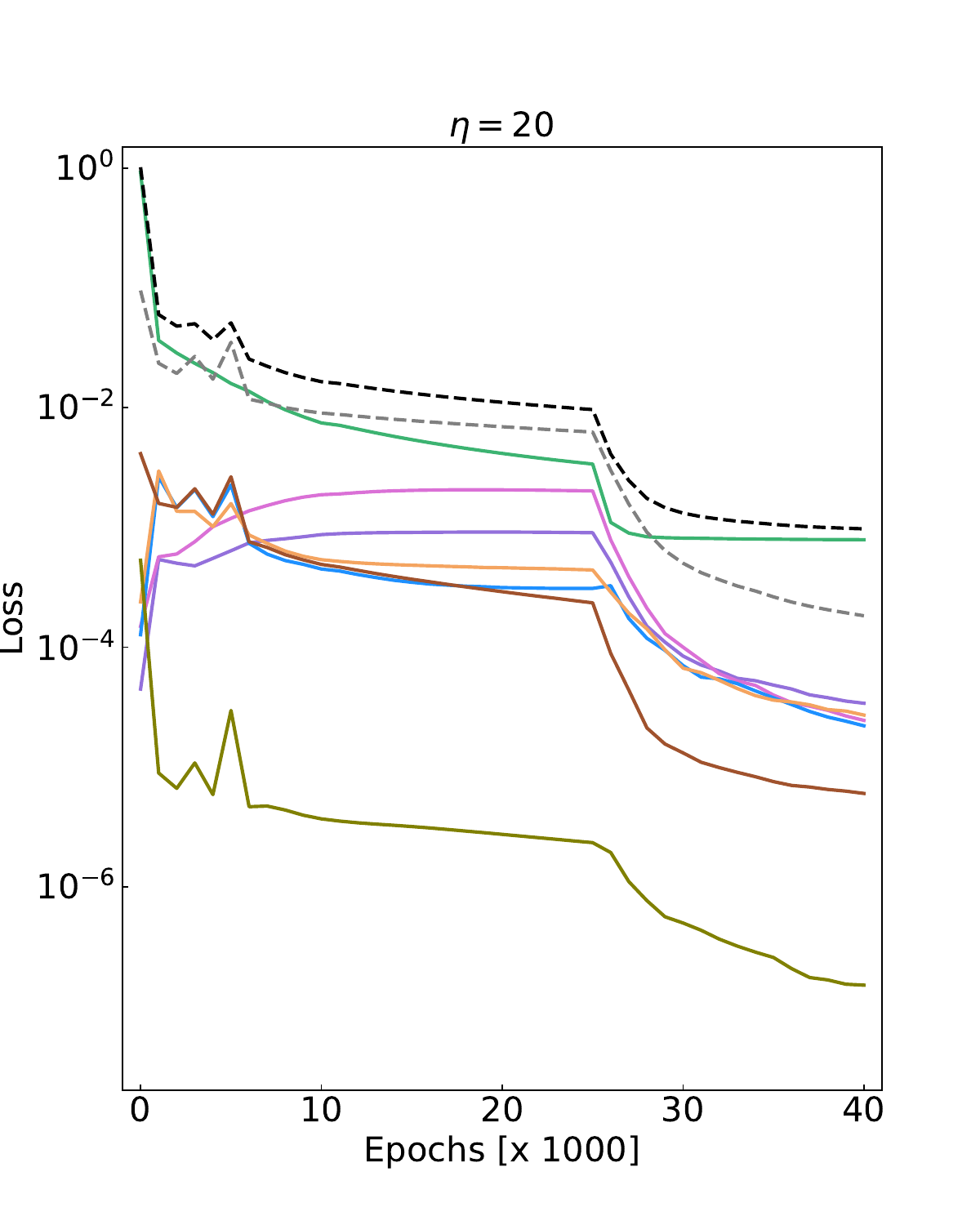}};

        \draw[ultra thick, dashed, black] (-3,-5.8)--(-2,-5.8);
        \draw[black](-0.4,-5.7) node {$\sum_{i = 1}^{15}{\lambda_i}{\mathcal{L}_i}
        + {\mathcal{L}_{\text{energy}}}$};

        \draw[ultra thick, dashed, gray] (2.2,-5.8)--(3.2,-5.8);
        \draw[black](4,-5.8) node {$\sum_{i = 1}^{15}{\lambda_i}{\mathcal{L}_i}$};

        \draw[ultra thick, sienna] (6,-5.8)--(7,-5.8);
        \draw[black](7.7,-5.8) node {$\sum_{i = 6}^{15}\mathcal{L}_{i}$};

        \draw[ultra thick, mediumseagreen] (9.6,-5.8)--(10.6,-5.8);
        \draw[black](11.3,-5.8) node {$\mathcal{L}_{\text{energy}}$};

        \draw[ultra thick, mediumpurple] (-3,-6.5)--(-2,-6.5);
        \draw[black](-1.6,-6.5) node {$\mathcal{L}_{1}$};

        \draw[ultra thick, orchid] (0.15,-6.5)--(1.15,-6.5);
        \draw[black](1.55,-6.5) node {$\mathcal{L}_{2}$};

        \draw[ultra thick, dodgerblue] (3.3,-6.5)--(4.3,-6.5);
        \draw[black](4.7,-6.5) node {$\mathcal{L}_3$};

        \draw[ultra thick, sandybrown] (6.45,-6.5)--(7.45,-6.5);
        \draw[black](7.85,-6.5) node {$\mathcal{L}_4$};

        \draw[ultra thick, olive] (9.6,-6.5)--(10.6,-6.5);
        \draw[black](11,-6.5) node {$\mathcal{L}_{5}$};
\end{tikzpicture}
\caption{The loss of the example B during the PINN training for two different values of 
$\eta$, i.e., $\eta = 1$ (left), and $\eta = 20$ (right).  
The black dashed line indicates the total loss $\mathcal{L}_{\text{2D}}(\Theta)$. 
The gray dashed line indicates the total loss minus the energy term 
(i.e., $\mathcal{L}_{\text{energy}}$, indicated by the green solid line). 
The line labeled $\sum_{i = 6}^{15}\mathcal{L}_{i}$ indicates the sum of loss terms  
associated with boundary conditions. The lines labeled $\mathcal{L}_i, 
i \in \{1,2,\cdots, 5\}$ indicate the loss derived from the governing equations 
Eqs. (\ref{eq_TIGL})(\ref{eq_momentum})(\ref{eq_Maxwell}).
All lines in this figure are plotted using the loss data evaluated every 1000 epochs.}
\label{fig_loss_B}
\end{figure}

Now we discuss the influence of the parameter $\eta$ in the energy term 
Eq. (\ref{eq_loss_total_energy}) on the PINN training. Two cases are considered:
$\eta = 1$ and $\eta = 20$. When $\eta = 1$, the relative errors are 
$\mathcal{E}(\tilde{P}_1)$ = 7.20$\%$, $\mathcal{E}(\tilde{P}_2)$ = 1.89$\%$, 
a bit worse than the PINN prediction of $\eta = 20$. This can be explained 
by the training loss, which is shown in Fig. \ref{fig_loss_B}. 
The reference value of attainable minimum energy for the example B is -0.351 aJ, 
so ideally we have $\min(\mathcal{L}_{\text{energy}})$ = 7.05 $\times$ 10$^{-1}$ 
if $\eta = 1$, and $\min(\mathcal{L}_{\text{energy}})$ = 9.12 $\times$ 10$^{-4}$
if $\eta = 20$. In the PINN implementation, we obtain 6.93 $\times$ 10$^{-1}$ 
(i.e., $\Psi \approx -0.367$) and 7.90 $\times$ 10$^{-4}$ (i.e., $\Psi \approx -0.357$)
for $\mathcal{L}_{\text{energy}}$ when $\eta = 1$ and $\eta = 20$, respectively, 
at the end of the training. Therefore, a good approximation of the minimum 
total free energy is achieved for either case. However, if the expected value of 
$\mathcal{L}_{\text{energy}}$ is too large, namely close to 1, 
the training could be trapped in a ``lazy" state where the PINN predicts 
a trivial solution such that $\mathcal{L}_{\text{energy}} \approx 1$, 
and other loss terms are extremely small. This can be clearly seen in the left 
of Fig. \ref{fig_loss_B}. The escape of this trap may require 
a lot of epochs or fail, resulting in an inefficient or unsuccessful training.
Besides, a too large $\mathcal{L}_{\text{energy}}$ probably hampers 
the decrease of other loss terms because the optimizer tends to focus on the  
largest loss term. This is likely the reason why the total loss excluding 
$\mathcal{L}_{\text{energy}}$ (i.e., the gray dashed lines in Fig. \ref{fig_loss_B}) 
of $\eta = 1$ is larger than that of $\eta = 20$. On the other hand, 
if choosing a large $\eta$ to make $\mathcal{L}_{\text{energy}}$ too small, 
the energy term will be inconsequential in the total loss Eq. (\ref{eq_loss_2D}), 
which is apparently not favorable. To sum up, the value of $\mathcal{L}_{\text{energy}}$ 
should be limited to a suitable range to ensure that the PINN training works properly. 
This limitation can be realized by adjusting the parameter $\eta$. 
As a rough guideline, a small $\eta$, e.g., $\eta = 1$ may serve as an initial try 
at the implementation to estimate the value of attainable minimum energy, 
and based on that, a good $\eta$  could be computed to make 
$\mathcal{L}_{\text{energy}}$ fall into a suitable range, e.g., 
between $10^{-3}$ and  $10^{-4}$. 
 
\subsection{One example in 3D}
This example is the 3D case of the example A in Section \ref{sect_2D_example}. 
The 3D computational domain $\Omega$ is a 4 nm $\times$ 4 nm $\times$ 0.4 nm 
thin plate in a Cartesian coordinate system $O-{x_1}{x_2}{x_3}$. The origin $O$ 
is located at the center of the plate, thus $x_1 \in [-2, 2], \: x_2 \in  [-2, 2], 
\: x_3 \in [-0.2, 0.2]$. The network for the 3D case has 3 nodes 
in the input layer for $x_1, x_2, x_3$, and 7 nodes in the output layer 
for $\tilde{P}_1, \tilde{P}_2, \tilde{P}_3, \tilde{u}_1, \tilde{u}_2, \tilde{u}_3, \tilde{\phi}$,  
as shown in Fig. \ref{fig_overview_3D}.

\subsubsection{Loss function}
We denote the boundaries satisfying $x_1 = \pm 2$ as $\Gamma_1$, 
$x_2 = \pm 2$ as $\Gamma_2$, and $x_3 = \pm 0.2$ as $\Gamma_3$. 
The loss function for the 3D example based on (\ref{eq_loss_pde})
(\ref{eq_loss_bc})(\ref{eq_loss_total_energy}) reads 
\begin{align}
        \mathcal{L}_{\text{3D}}(\Theta) = \sum_{i = 1}^{25}{\lambda_i}{\mathcal{L}_i}
        + {\mathcal{L}_{\text{energy}}}, 
        \label{eq_loss_3D}
\end{align}
where 
\begin{subequations}
        \begin{align}
                &\mathcal{L}_i =\bigg |\frac{\partial{\tilde{\psi}}}{\partial{\tilde{{P}}_i}} 
                                - \left(\frac{\partial}{\partial{x_1}}(\frac{\partial{\tilde{\psi}}}{\partial{\tilde{{P}}_{i,1}}})
                                      + \frac{\partial}{\partial{x_2}}(\frac{\partial{\tilde{\psi}}}{\partial{\tilde{{P}}_{i,2}}})
                                      + \frac{\partial}{\partial{x_3}}(\frac{\partial{\tilde{\psi}}}{\partial{\tilde{{P}}_{i,3}}})\right)\bigg |_{\Omega}^2, 
                                \quad i \in \{1, 2, 3\}, \label{eq_loss_TIGL_3D}\\
               &\mathcal{L}_i = |\tilde{\sigma}_{j1,1} + \tilde{\sigma}_{j2,2} + \tilde{\sigma}_{j3,3}|_{\Omega}^2,  
                                \quad j = i - 3, \quad i \in \{4, 5, 6\},  \label{eq_loss_mechanic_3D} \\
               &\mathcal{L}_7 = |\tilde{D}_{1,1} + \tilde{D}_{2,2} + \tilde{D}_{3,3} |_{\Omega}^2,  
                                \label{eq_loss_Maxwell_3D} \\
                \begin{split}
                        &\mathcal{L}_i =\bigg |\frac{\partial{\tilde{\psi}}}{\partial{\tilde{{P}}_{j,1}}}\bigg |_{\Gamma_{1}}^2,  
                                        \: j = i - 7, \: i \in \{8, 9, 10\},    \quad 
                         \mathcal{L}_i =\bigg |\frac{\partial{\tilde{\psi}}}{\partial{\tilde{{P}}_{j,2}}}\bigg |_{\Gamma_{2}}^2, 
                                        \: j = i - 10, \: i \in \{11, 12, 13\},  \\
                        &\mathcal{L}_i =\bigg |\frac{\partial{\tilde{\psi}}}{\partial{\tilde{{P}}_{j,3}}}\bigg |_{\Gamma_{3}}^2, 
                                        \: j = i - 13, \: i \in \{14, 15, 16\},                            
                \label{eq_loss_grad_3D}
                \end{split}
                \\
                        &\mathcal{L}_{17} = |\tilde{\sigma}_{11} |_{\Gamma_1}^2, \:
                         \mathcal{L}_{18} = |\tilde{\sigma}_{22} |_{\Gamma_2}^2, \:
                         \mathcal{L}_{19} = |\tilde{\sigma}_{33} |_{\Gamma_3}^2, \:
                         \mathcal{L}_{20} = |\tilde{\sigma}_{12} |_{\Gamma_1 \cup \Gamma_2}^2, \:
                         \mathcal{L}_{21} = |\tilde{\sigma}_{23} |_{\Gamma_2 \cup \Gamma_3}^2, \:
                         \mathcal{L}_{22} = |\tilde{\sigma}_{13} |_{\Gamma_1 \cup \Gamma_3}^2, 
                 \label{eq_loss_traction_3D}
                \\
                 &\mathcal{L}_{23} = |\tilde{D}_1 |_{\Gamma_1}^2, \quad
                  \mathcal{L}_{24} = |\tilde{D}_2 |_{\Gamma_2}^2, \quad
                  \mathcal{L}_{25} = |\tilde{D}_3 |_{\Gamma_3}^2. 
                \label{eq_loss_electric_3D}
        \end{align}
        \label{eq_loss_terms_3D}
 \end{subequations}
For this 3D example, we still choose $\Psi^0 = 0$ when computing 
$\mathcal{L}_{\text{energy}}$ in Eq. (\ref{eq_loss_3D}).

\subsubsection{Results}
For the 3D example, we use a network with 3 hidden layers and 40 nodes in each hidden layer 
to predict the ferroelectric microstructure at equilibrium.
The values of the 25 weights $\lambda_i$ in the loss function Eq. (\ref{eq_loss_3D}) 
are: $\lambda_i = 1$ when $i \in \{1,2, \cdots, 6\}$, $\lambda_7 = 100$, and 
$\lambda_i = 10$ when $i \in \{8, 9, \cdots, 25\}$. 
The parameter $\eta$ is set to be 100. This PINN is trained on 46000 collocation points 
for 100000 epochs, which takes about 10 hours.

\begin{figure}[H]
        \centering 
        { 
        \includegraphics[scale=0.35]{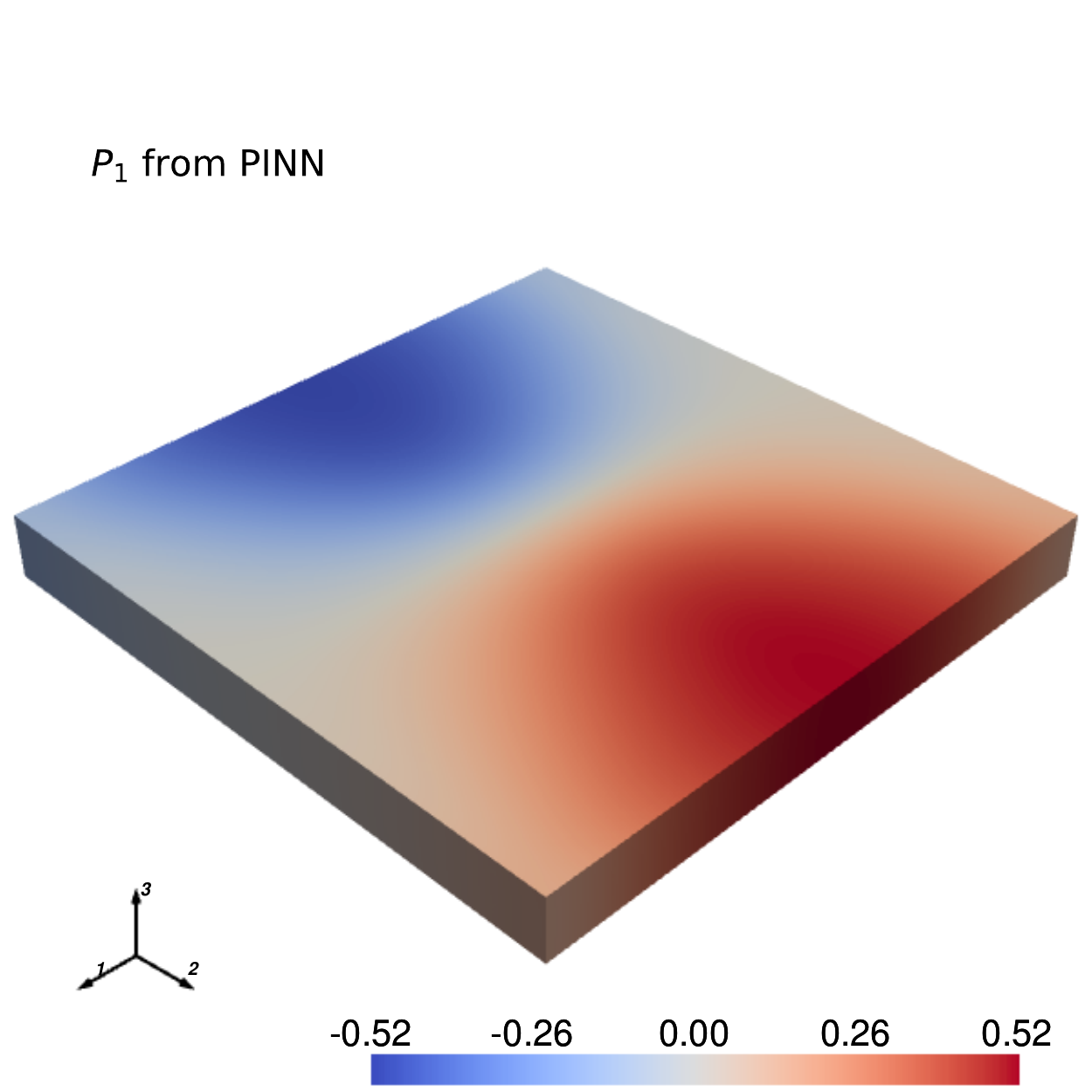} 
        \hspace{1.4cm}
        \includegraphics[scale=0.35]{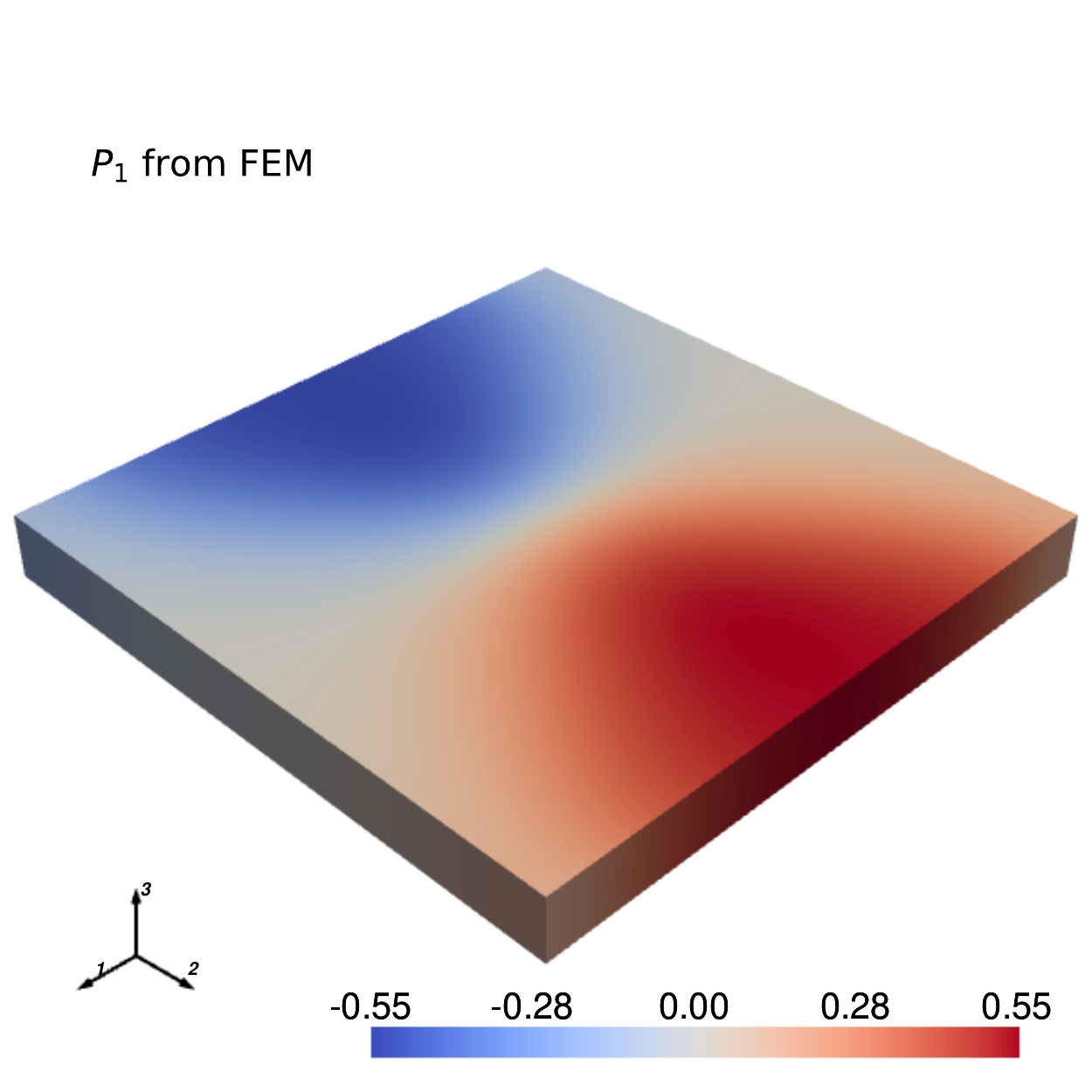}
        }
        \caption{The polarization component $P_1$ at equilibrium obtained from 
        the PINN (left) and FEM (right) for the 3D example. 
        Here the PINN figure is plotted using the opposite numbers of the raw PINN data.}
\label{fig_3D_P1}
\end{figure}

\begin{figure}[H]
        \centering 
        { 
        \includegraphics[scale=0.35]{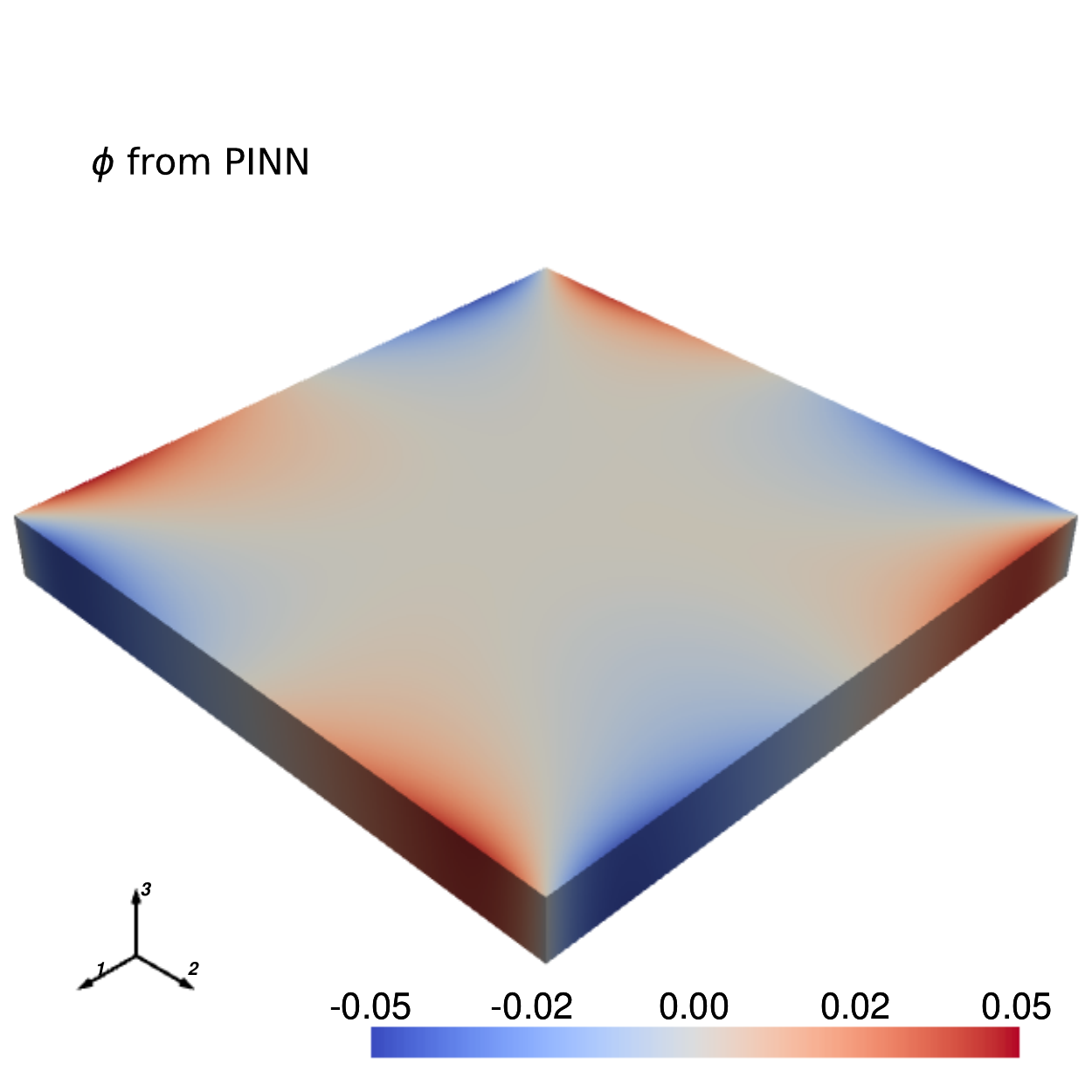} 
        \hspace{1.4cm}
        \includegraphics[scale=0.35]{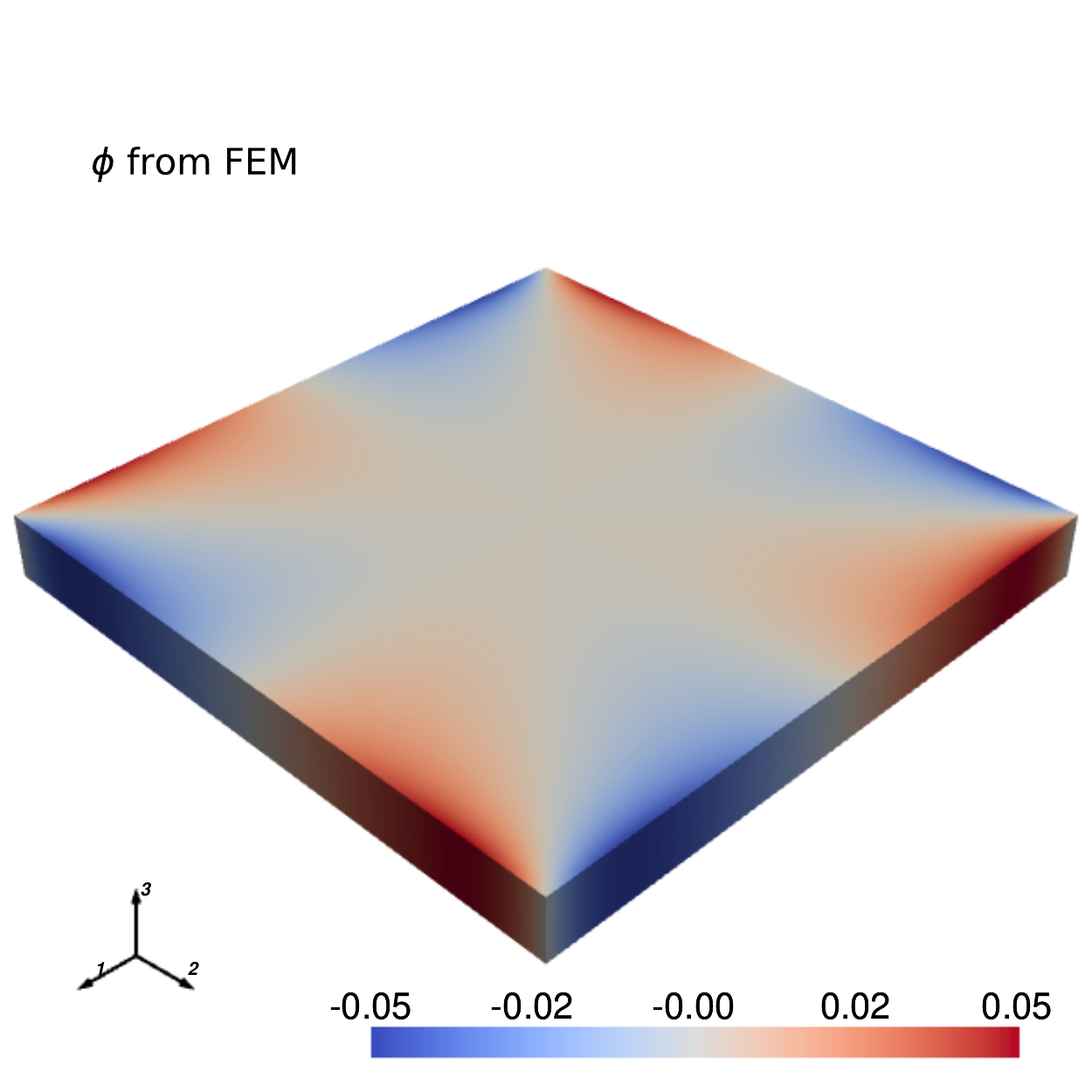}
        }
        \caption{The electric potential $\phi$ at equilibrium obtained from 
        the PINN (left) and FEM (right) for the 3D example. 
        Here the PINN figure is plotted using the opposite numbers of the raw PINN data.}
\label{fig_3D_phi}
\end{figure}

The results are given partly to save space in Fig. \ref{fig_3D_P1} and Fig. \ref{fig_3D_phi}, 
which show the PINN versus FEM comparison of the polarization component $P_1$ 
and the electric potential $\phi$, respectively.
The visualization of $P_2$ is similar to that of $P_1$ (see 
Fig. \ref{fig_P1P2_A} for example), so the comparison of $P_2$
is not present here. The relative errors of $P_1, \: P_2$ are 
$\mathcal{E}(\tilde{P}_1)$ = 10.2$\%$, \: $\mathcal{E}(\tilde{P}_2)$ = 10.1$\%$.
As for $P_3$, since its magnitude over such a thin plate is 
extremely small ($\approx 0$), the visualization would be trivial, 
and Eq. (\ref{eq_relative_error}) would not be a good indicator  
to evaluate the accuracy of $\tilde{P}_3$. Instead, we ascertain 
the PINN prediction of $P_3$ is reasonable via the examination that 
the maximum absolute value of $\tilde{P}_3$ 
($\max(|\tilde{P}_3|) = 5.98 \times 10^{-3}$) is close to 0. 
Given the above, 
it is fair to say that the en-PF PINN approach can effectively 
predict the steady ferroelectric microstructure for the 3D example.

\section{The en-PF PINN approach helps to solve a dynamic problem}
\label{sect_dynamic_example}
In Section \ref{sect_examples}, the en-PF PINN approach 
(as seen in Fig. \ref{fig_overview_3D})
is used to predict the steady ferroelectric microstructure
without tracking the evolution process. Hence, Section \ref{sect_examples} 
actually deals with static problems. By contrast, the current  
section focuses on a dynamic problem. Although the en-PF PINN 
approach cannot solve a dynamic problem, it provides  
useful data that make a data-guided PINN possible. In what follows, 
the superiority of the data-guided PINN over the baseline PINN 
for a dynamic problem of ferroelectric microstructure evolution 
is demonstrated.

\subsection{Problem setup}
We consider the dynamic case of the example A 
described in Section \ref{sect_2D_example}. The initial condition  
is prescribed as  
\begin{align}
        P_1^{\text{init}} = P_2^{\text{init}} = \cos(0.5{\pi}{x_1})\sin(0.25{\pi}{x_2}).
        \label{eq_prescribed_ic}
\end{align}
We choose Eq. (\ref{eq_prescribed_ic}) instead of a random normal distribution 
because it takes less time for the ferroelectric system to achieve the equilibrium 
with such an initial condition. A shorter evolution process will 
naturally make the associated PINN training easier. With Eq. (\ref{eq_prescribed_ic}), 
the whole evolution process takes about 16 ps, which is assumed to be known 
for the data-guided PINN.

\subsection{Loss function}
For the dynamic problem Eqs. (\ref{eq_TDGL})-(\ref{eq_Maxwell}) 
along with the boundary conditions indicated in Fig. \ref{fig_2d_bcs} 
(for the example A) and the initial condition Eq. (\ref{eq_prescribed_ic}), 
the loss function of a baseline PINN is similar to Eq. (\ref{eq_loss_2D}), 
but the energy term $\mathcal{L}_{\text{energy}}$ should not be included, 
and the loss terms in Eq. (\ref{eq_loss_TIGL_2D}) need to be modified 
based on Eq. (\ref{eq_TDGL}). The initial condition Eq. (\ref{eq_prescribed_ic}) 
is enforced as a hard constraint, so there are no loss terms 
associated with the initial condition. The loss function of the data-guided PINN 
contains all terms present in the loss function of the baseline PINN, 
and additionally, a loss term to measure the mismatch between the network  
outputs $\tilde{P}_1, \tilde{P}_2$ and the polarizations obtained beforehand 
in Section \ref{sect_result_example_A} is required. 
Denoting the known polarizations as $\bar{P}_1, \bar{P}_2$, the additional 
loss term of the data-guided PINN can be written as 
\begin{align}
        \mathcal{L}_{\text{data}} = \sum_i^{N_{\text{d}}}(|\tilde{P}_1(\bm{x}_i, t = 16) 
                                   - \bar{P}_1(\bm{x}_i, t = 16)|^2 
                                   +|\tilde{P}_2(\bm{x}_i, t = 16) - \bar{P}_2(\bm{x}_i, t = 16)|^2),
                                   \label{eq_loss_data}
\end{align}
where $N_{\text{d}}$ is the number of labeled data 
$\bar{P}_1(\bm{x}_i, t =16)$ or $\bar{P}_2(\bm{x}_i, t =16)$.

\subsection{Results}
We solve the above dynamic problem of ferroelectric microstructure evolution 
using the data-guided PINN or the baseline PINN, with the same 
network architecture: 3 nodes in the input layer, 5 nodes in the output layer, 
and 3 hidden layers, 80 nodes for each. The weights for the loss terms are: 
100 for terms derived from Eq. (\ref{eq_TDGL}), 
1 for terms derived from Eqs. (\ref{eq_momentum})(\ref{eq_Maxwell}), 
10 for terms associated with the boundary conditions, 
and 1 for $\mathcal{L}_{\text{data}}$ in the data-guided PINN.  
In the implementation of the baseline PINN, we simply set the weight 
for $\mathcal{L}_{\text{data}}$ to be 0, and there are no other modifications 
to the code of the data-guided PINN. Since the network size is much larger 
than that of the example A in Section \ref{sect_2D_example} 
and more epochs are necessary, the PINN training is much more expensive, 
taking almost 5 hours.

\begin{figure}[H]
        \centering 
        { 
        \includegraphics[scale=0.35]{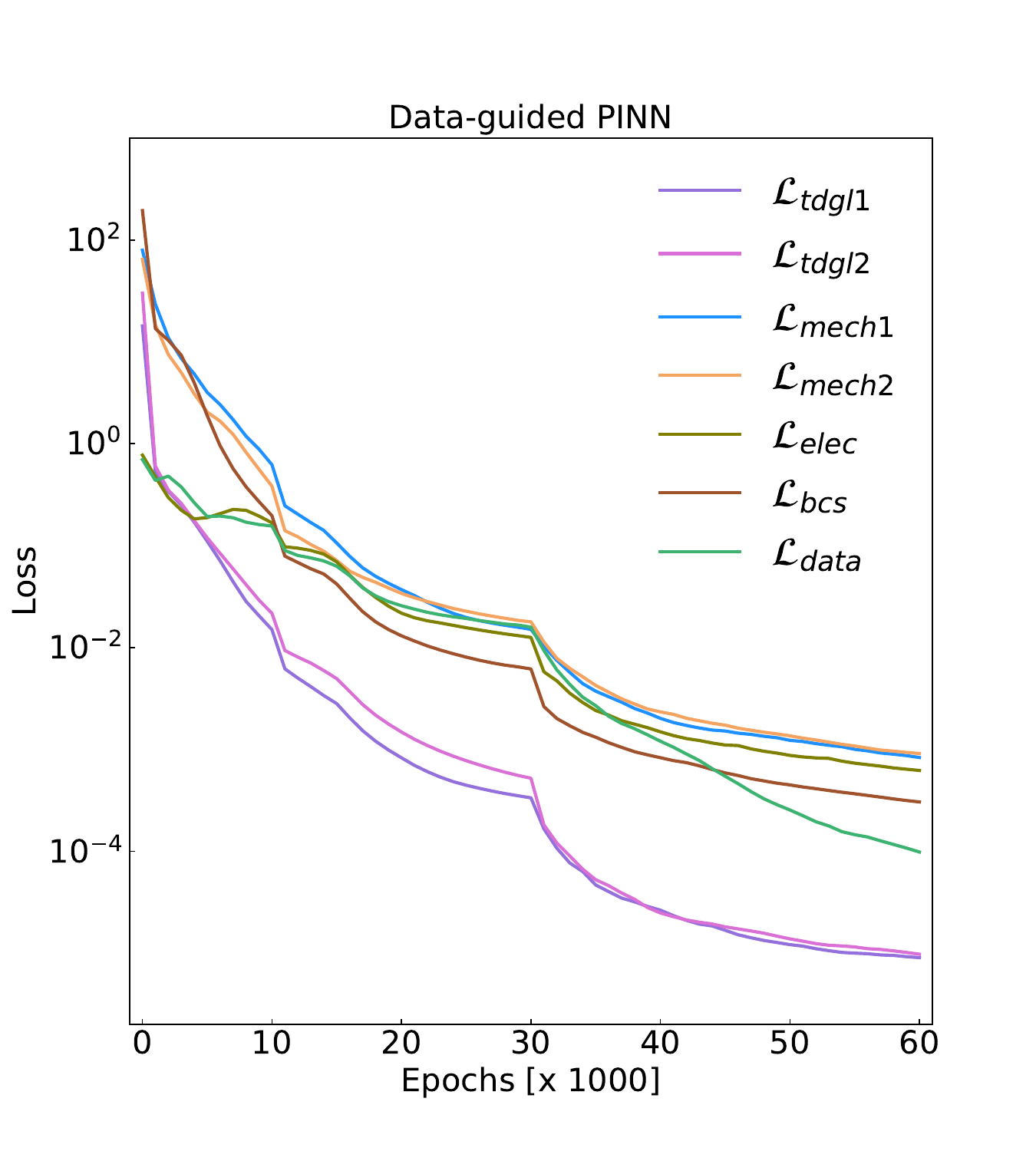} 
        \includegraphics[scale=0.35]{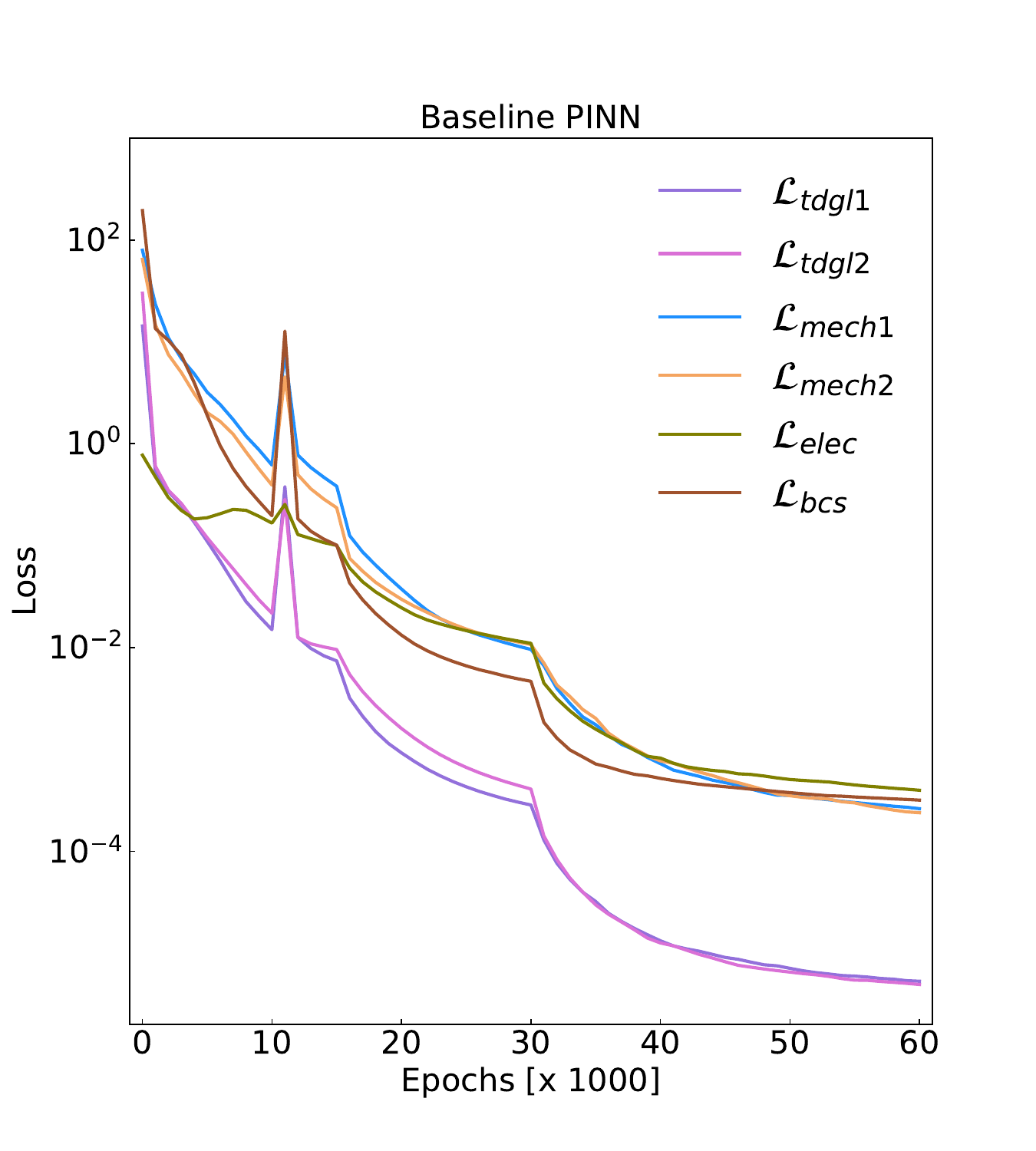}
        }
        \caption{The loss of data-guided PINN (left) and the baseline PINN (right). 
        $\mathcal{L}_{\text{tdgl1}}, \mathcal{L}_{\text{tdgl2}}$ indicate 
        the loss terms derived from the TDGL equation (i.e., Eq. (\ref{eq_TDGL})), 
        $\mathcal{L}_{\text{mech1}}, \mathcal{L}_{\text{mech2}}$ indicate 
        the loss terms derived from the equation of momentum balance 
        (i.e., Eq. (\ref{eq_momentum})), $\mathcal{L}_{\text{elec}}$ indicates 
        the loss term derived from the Maxwell's equation (i.e., Eq. (\ref{eq_Maxwell})), 
        and $\mathcal{L}_{\text{bcs}}$ indicates the sum of loss terms 
        associated with boundary conditions. Besides, $\mathcal{L}_{\text{data}}$ 
        in the left figure indicates the loss computed from the data 
        predicted beforehand by the en-PF PINN approach. 
        }
\label{fig_loss_dynamic_case}
\end{figure}

\begin{figure}[H]
        \centering 
        { 
        \includegraphics[scale=0.35]{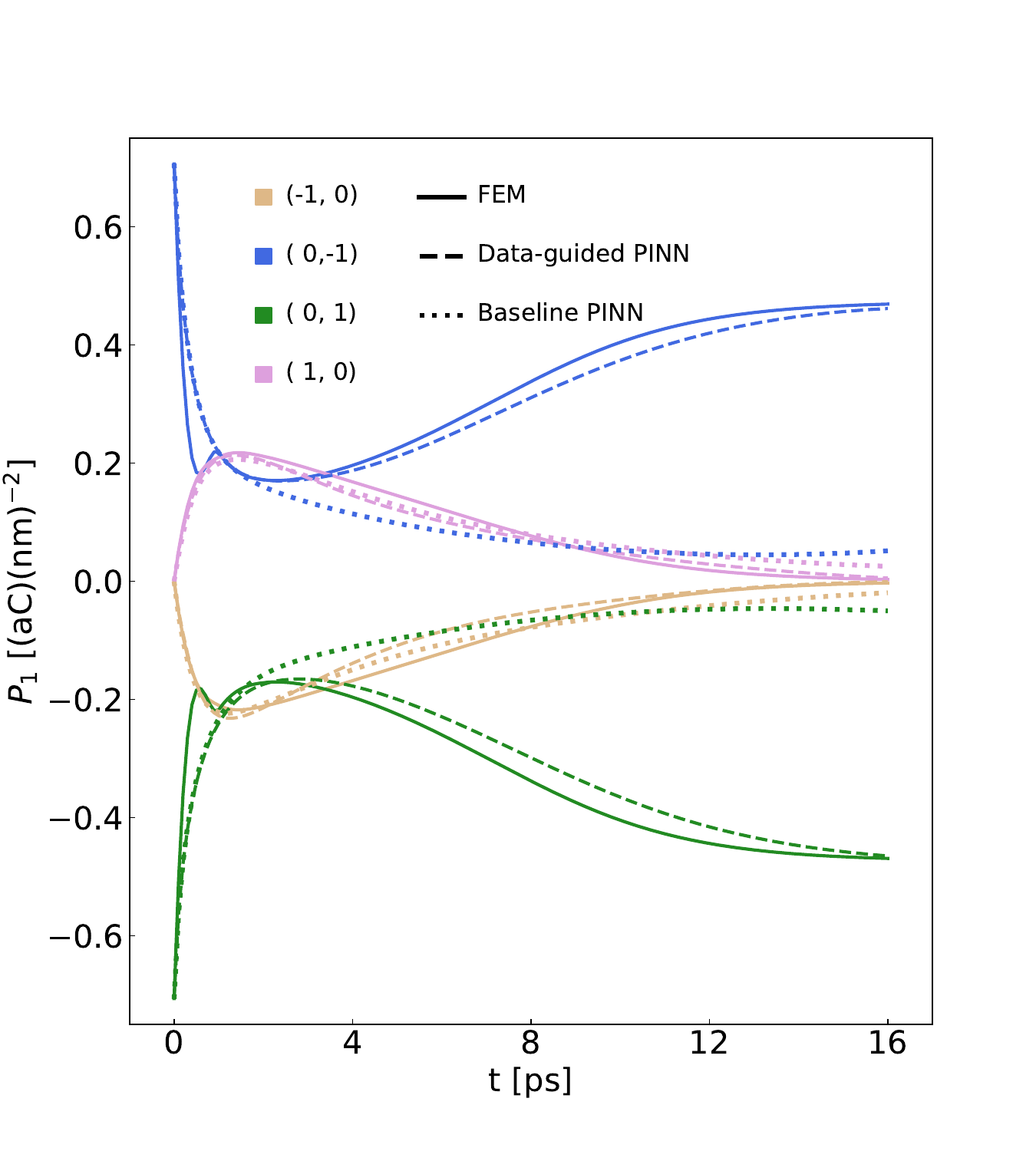}  
        \includegraphics[scale=0.35]{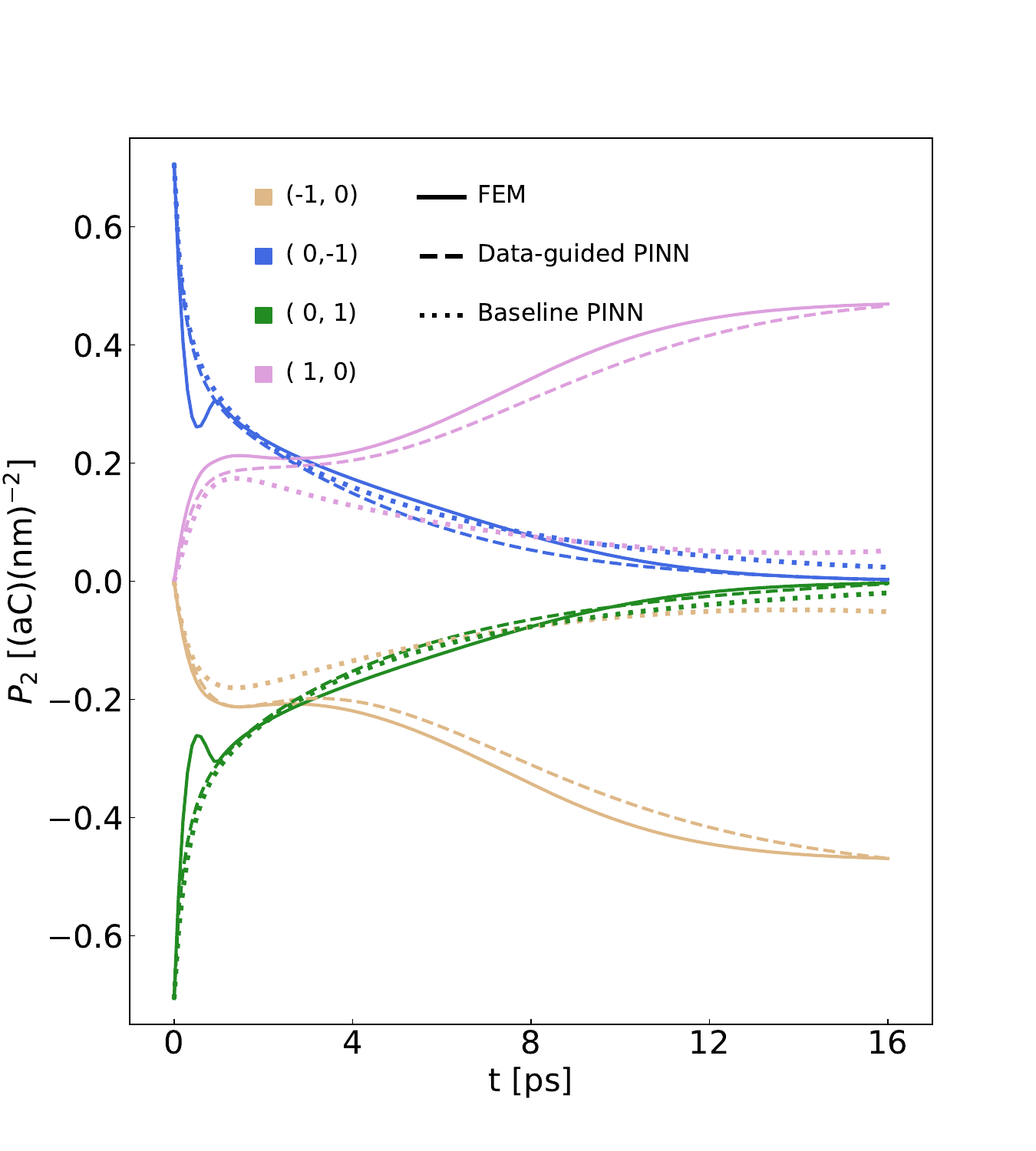}
        }
        \caption{Evolution of the polarization components $P_1$ (left) and 
        $P_2$ (right) at four locations ($\pm$ 1, 0), (0, $\pm$ 1) predicted 
        by the data-guided PINN or the baseline PINN.}
\label{fig_polarization_evolution}
\end{figure}

The loss during the training course is shown in Fig. \ref{fig_loss_dynamic_case}, 
where we see that the baseline PINN has nearly the same or even smaller 
values of different loss terms than the data-guided PINN. However, 
the baseline PINN does not predict better results.
To visualize this, Fig. \ref{fig_polarization_evolution} gives 
the polarization evolution at four locations predicted by the two PINNs. 
It is clear from Fig. \ref{fig_polarization_evolution} that the 
data-guided PINN manages to produce results close to the reference FEM solution, 
while the baseline PINN does the opposite. In the initial stage,
the solution predicted by the baseline PINN agrees well with the reference solution,  
but as the evolution progresses, the baseline PINN seems to encounter the 
propagation failure \cite{daw2022mitigating}, getting stuck at a trivial solution. 
Conversely, with the help of the labeled data obtained beforehand, 
the data-guided PINN eludes this failure, and predicts a good solution 
over the whole evolution process. This comparison shows that the en-PF PINN approach 
also plays a positive role in the dynamic case.

\paragraph{Remarks} 
{It is noteworthy that the prediction of long-term dynamic behaviors  
is usually challenging for PINNs. 
One reason is the propagation failure \cite{daw2022mitigating}, 
which means that the propagation of the solution from the boundary 
and/or initial conditions to interior points is somehow disrupted, 
thus leading to unexpected solutions. More detailed explanations of 
this failure mode can be found in \cite{rohrhofer2022role}. 
To mitigate this issue, some remedies such as residual-based 
resampling \cite{daw2022mitigating} and causal training  
\cite{mattey2022novel, wang2022respecting} have been proposed. 
In the current work, we present a simple method to overcome this failure, 
i.e., the introduction of labeled data at equilibrium into the PINN loss function. 
The key to this method is that the labeled data can be easily obtained 
from the en-PF PINN approach, as shown in Section \ref{sect_examples}. 
From this point of view, this method is obviously not general, 
but it is effective on the specific dynamic problem 
of ferroelectric microstructure evolution.}

\section{Conclusions}
\label{sect_conclusions}
This work develops a PINN approach for the prediction of steady 
ferroelectric microstructures. In simulations, ferroelectric microstructures 
are usually obtained by solving a phase-field model, which is  
a time-dependent, nonlinear, and high-order PDE system of multi-physics.   
It is difficult to solve such a PDE system with a baseline PINN. 
For ease of handling, we reformulate the PDE system to be a static problem 
instead of solving it directly. This reformulation is physically valid, 
but the resulting static problem is not well-posed in mathematics. 
Hence, we introduce an extra constraint inspired by
the law of energy dissipation into the loss function of a PINN to 
achieve the expected solution. This PINN approach is validated 
against two 2D examples and one 3D example, 
where the PINN results of the steady ferroelectric 
microstructure are compared with the FEM results.

The above examples show that the proposed PINN approach 
is capable of capturing steady ferroelectric microstructures, 
which is a primary goal in simulations of ferroelectric 
microstructure evolution. Besides, if the evolution process is 
of interest, the proposed PINN approach remains helpful because 
it can provide labeled data. These data are crucial to the 
successful PINN prediction of the complete process of ferroelectric 
microstructure evolution. The reason is that the labeled data 
are able to assist the PINN of the dynamic problem in preventing  
the propagation failure, a common failure mode faced with PINNs 
when predicting dynamic behaviors. We demonstrate this through 
a 2D example concerning the evolution process.

The proposed PINN approach undoubtedly accelerates the prediction 
of steady ferroelectric microstructures. However, this statement 
is true only in the PINN framework. If compared with FEM, the 
proposed PINN approach is still computationally expensive.
Therefore, further study of how to reduce the PINN computational 
burden is significant, and that could be one topic of our future research.

\section*{CODE AVAILABLITY}
The source code used in this work can be found at 
\href{https://github.com/whshangl/Open-en-PF-PINN}{Open en-PF PINN}.

\section*{ACKNOWLEDGMENTS}
The authors gratefully acknowledge the support of 
the National Program on Key Basic Research Project (Grant No.2022YFB3807601), 
National Natural Science Foundation of China (Grant No. 12272338 and 12192214),
and China Postdoctoral Science Foundation (Grant No. 2023TQ0332).

\section*{Appendix A}
\label{sect_appendx_A}
For the microstructure evolution of a ferroelectric monocrystal, 
the total free energy is composed of the Landau energy, 
the gradient energy, the strain energy, and the electrostatic energy. 
Thus, the free energy density $\psi$ reads
\begin{align}
        \psi = \psi_{\text{Landau}} + \psi_{\text{gradient}} + \psi_{\text{strain}} + \psi_{\text{electrostatic}}, 
        \label{eq_free_energy}
\end{align}
where the components in $\psi$ are \cite{li2002effect, su2007continuum}
\begin{subequations}
        \begin{align}
                \begin{split}
                        \psi_{\text{Landau}} = 
                        &+ \alpha_1(P_1^{2} + P_2^{2}+P_3^{2}) 
                         + \alpha_{11}(P_1^{4} + P_2^{4}+P_3^{4})
                         + \alpha_{12}(P_1^{2}P_2^{2}+P_2^{2}P_3^{2}+P_1^{2}P_3^{2}) \\
                        &+ \alpha_{111}(P_1^{6} + P_2^{6}+P_3^{6})
                         + \alpha_{123}P_1^{2} P_2^{2} P_3^{2}\\
                        &+ \alpha_{112}\left(P_1^{4}(P_3^{2}+P_2^{2})+P_2^{4}(P_1^{2}+P_3^{2})+P_3^{4}(P_1^{2}+P_2^{2})\right),
                \label{eq_Landau_energy}
                \end{split} 
                \\
                \begin{split}
                        \psi_{\text{gradient}} = 
                      & +\frac{1}{2}G_{11}(P_{1,1}^2 + P_{2,2}^2 + P_{3,3}^2)
                        + G_{12}(P_{1,1}P_{2,2} + P_{2,2}P_{3,3} + P_{1,1}P_{3,3}) \\
                      & +\frac{1}{2}G_{44}\left((P_{1,2} + P_{2,1})^2+(P_{1,3} + P_{3,1})^2 + (P_{2,3} + P_{3,2})^2\right) \\
                      & +\frac{1}{2}G_{44}^{\prime}\left((P_{1,2} - P_{2,1})^2+(P_{1,3} - P_{3,1})^2 + (P_{2,3} - P_{3,2})^2\right), 
                \label{eq_gradient_energy}
                \end{split}  
                \\
                \begin{split}
                        \psi_{\text{strain}} = 
                      & +\frac{1}{2}c_{11}(\varepsilon_{11}^2+\varepsilon_{22}^2+\varepsilon_{33}^2)
                        +c_{12}\left(\varepsilon_{11}\varepsilon_{22}+\varepsilon_{22}\varepsilon_{33}+\varepsilon_{11}\varepsilon_{33}\right)
                        +2c_{44}(\varepsilon_{12}^2+\varepsilon_{23}^2+\varepsilon_{13}^2) \\
                      & -q_{11}(\varepsilon_{11}P_1^{2}+\varepsilon_{22}P_2^{2}+\varepsilon_{33}P_3^{2})
                        -2q_{44}(\varepsilon_{12}{P_1}{P_2} + \varepsilon_{13}{P_1}{P_3} + \varepsilon_{23}{P_2}{P_3})\\
                      & -q_{12}\left(\varepsilon_{11}(P_3^{2}+P_2^{2})+\varepsilon_{22}(P_1^{2}+P_3^{2})+\varepsilon_{33}(P_1^{2}+P_2^{2})\right) \\
                      & + \beta_{11}(P_1^{4} + P_2^{4}+P_3^{4}) + \beta_{12}(P_1^{2}P_2^{2}+P_2^{2}P_3^{2}+P_1^{2}P_3^{2}),
                \label{eq_strain_energy}
                \end{split}
                \\
                \begin{split}
                        \psi_{\text{electrostatic}} = 
                        &-\frac{1}{2}{\kappa}_c\left(E_1^2 + E_2^2 + E_3^2\right) -E_1P_1-E_2P_2-E_3P_3.
                \label{eq_electrostatic_energy}
                \end{split}
        \end{align}
\end{subequations}   
In $\psi_{\text{Landau}}$, $\alpha_1, \alpha_{11}, \alpha_{12}, \alpha_{111}, \alpha_{112}, \alpha_{123}$ 
are the expansion coefficients. 
In $\psi_{\text{gradient}}$, $G_{11}, G_{12}, G_{44}, G_{44}^{\prime}$ are gradient energy coefficients.
In $\psi_{\text{strain}}$, $c_{11}, c_{12}, c_{44}$ are elastic constants,
$q_{11}, q_{12}, q_{44}$ are electrostrictive coefficients, and the last line  
is related to the stress-free strain caused by the polarization field \cite{li2002effect}.
In $\psi_{\text{electrostatic}}$ \footnote{The electrostatic energy density 
should be $\frac{1}{2}{{\kappa}_c}(E_1^2 + E_2^2 + E_3^2)$ in the strict sense. 
Eq. (\ref{eq_electrostatic_energy}) here is actually the term in the 
electric enthalpy derived from Legendre transformation to facilitate 
the constitutive relations concerned in the phase-field model of interest 
\cite{su2007continuum}. We can use either form of $\psi_{\text{electrostatic}}$ 
(i.e., Eq. (\ref{eq_electrostatic_energy}) or 
$\psi_{\text{electrostatic}} = \frac{1}{2}{{\kappa}_c}(E_1^2 + E_2^2 + E_3^2)$)
to compute the total free energy in Eq. (\ref{eq_loss_total_energy}).
}, 
$\kappa_c$ is the dielectric constant.

\section*{Appendix B}
\label{sect_appendx_B}
The weak form of the phase-field model Eqs. (\ref{eq_TDGL})-(\ref{eq_bc_electric}) is 
\begin{align}
        \begin{split}
        0 = &+\int_{\Omega}\left({\sigma_{ij}}{\delta{\varepsilon_{ij}}} - {D_i}{\delta{E_i}} 
             -(\frac{1}{\zeta}\frac{\partial{P}_i}{\partial{t}} + \frac{\partial{\psi}}{\partial{P}_i}){\delta{P}_i}
             -\frac{\partial{\psi}}{\partial{P}_{i,j}}{\delta{P}_{i,j}}\right)\text{d}{\bm{x}} \\
            &+\int_{\partial{\Omega}}(\bar{\pi}_i{\delta{P}_i} 
             -\bar{q}{\delta{\phi}}-\bar{\tau_i}{\delta{u}_i})\text{d}{\bm{s}}, 
        \end{split}
\end{align}
where the components $\varepsilon_{ij} (= \varepsilon_{ji})$, $\sigma_{ij} (= \sigma_{ji})$, 
$E_i$, $D_i$ are 
\begin{align}
         &\varepsilon_{11} = u_{1,1}, \; \varepsilon_{22} = u_{2,2}, \; \varepsilon_{33} = u_{3,3}, \;
          \varepsilon_{23} = \frac{1}{2}(u_{2,3} + u_{3,2}), \;
          \varepsilon_{13} = \frac{1}{2}(u_{1,3} + u_{3,1}), \;
          \varepsilon_{12} = \frac{1}{2}(u_{1,2} + u_{2,1}); \\
        \begin{split}
                &\sigma_{11} =  c_{11}{\varepsilon_{11}}+c_{12}({\varepsilon_{22}}+{\varepsilon_{33}})-q_{11}P_1^2-q_{12}(P_3^2+P_2^2), \\
                &\sigma_{22} =  c_{11}{\varepsilon_{22}}+c_{12}({\varepsilon_{11}}+{\varepsilon_{33}})-q_{11}P_2^2-q_{12}(P_1^2+P_3^2), \\
                &\sigma_{33} =  c_{11}{\varepsilon_{33}}+c_{12}({\varepsilon_{22}}+{\varepsilon_{11}})-q_{11}P_3^2-q_{12}(P_1^2+P_2^2), \\
                &\sigma_{23} =  2c_{44}{\varepsilon_{23}-q_{44}P_2P_3}, \quad
                \sigma_{13} =  2c_{44}{\varepsilon_{13}-q_{44}P_1P_3}, \quad
                \sigma_{12} =  2c_{44}{\varepsilon_{12}-q_{44}P_1P_2}; 
       \end{split}
       \\   
       &E_1 =  -\phi_{,1}, \quad E_2 = -\phi_{,2}, \quad E_3 = -\phi_{,3}; \\ 
       &D_1 = {\kappa}_c{E_1} + P_1, \quad  D_2 = {\kappa}_c{E_2} + P_2, \quad D_3 = {\kappa}_c{E_3} + P_3. 
\end{align} 
The expansion of $\frac{\partial{\psi}}{\partial{P_i}}$ is 
\begin{align}      
       \begin{split}
       \frac{\partial{\psi}}{\partial{P_1}}  
       =\: &  {+2{\alpha}_1P_1 + 4({\alpha}_{11} + \beta_{11})P_1^3 
              +2({\alpha}_{12} + {\beta}_{12})P_1(P_2^2 + P_3^2)} \\
          &  {+6{\alpha}_{111}P_1^5
             +4{\alpha}_{112}P_1^3(P_2^2+P_3^2) 
             +2{\alpha}_{112}P_1(P_2^4+P_3^4)
             +2{\alpha}_{123}P_1P_2^2P_3^2} \\
          &  {-2q_{11}{\varepsilon}_{11}P_1 
            -2q_{12}P_1({\varepsilon}_{22}+{\varepsilon}_{33})
            -2q_{44}({\varepsilon}_{12}P_2 + {\varepsilon}_{13}P_3)} 
            {-E_1}, 
       \\
        \frac{\partial{\psi}}{\partial{P_2}}  
        =\: &  {+2{\alpha}_1P_2 + 4({\alpha}_{11} + \beta_{11})P_2^3 
               +2({\alpha}_{12} + {\beta}_{12})P_2(P_1^2 + P_3^2)} \\
           &  {+6{\alpha}_{111}P_2^5
              +4{\alpha}_{112}P_2^3(P_1^2+P_3^2) 
              +2{\alpha}_{112}P_2(P_1^4+P_3^4)
              +2{\alpha}_{123}P_2P_1^2P_3^2} \\
           &  {-2q_{11}{\varepsilon}_{22}P_2 
             -2q_{12}P_2({\varepsilon}_{11}+{\varepsilon}_{33})
             -2q_{44}({\varepsilon}_{12}P_1 + {\varepsilon}_{23}P_3)} 
             {-E_2}, 
        \\
        \frac{\partial{\psi}}{\partial{P_3}}  
        =\: &  {+2{\alpha}_1P_3 + 4({\alpha}_{11} + \beta_{11})P_3^3 
                +2({\alpha}_{12} + {\beta}_{12})P_3(P_1^2 + P_2^2)} \\
        &  {+6{\alpha}_{111}P_3^5
                +4{\alpha}_{112}P_3^3(P_1^2+P_2^2) 
                +2{\alpha}_{112}P_3(P_1^4+P_2^4)
                +2{\alpha}_{123}P_3P_1^2P_2^2} \\
        &  {-2q_{11}{\varepsilon}_{33}P_3 
                -2q_{12}P_3({\varepsilon}_{11}+{\varepsilon}_{22})
                -2q_{44}({\varepsilon}_{13}P_1 + {\varepsilon}_{23}P_2)} 
                {-E_3}. 
        \end{split}
\end{align}
The expansion of $\frac{\partial{\psi}}{\partial{P}_{i,j}}$ is      
\begin{align}
        \begin{split}
          \frac{\partial{\psi}}{\partial{P}_{1,1}} 
        = & \: G_{11}{P}_{1,1} + G_{12}({P}_{2,2}+{P}_{3,3}), 
        \\
          \frac{\partial{\psi}}{\partial{P}_{2,2}} 
        = & \: G_{11}{P}_{2,2} + G_{12}({P}_{1,1}+{P}_{3,3}),
        \\ 
          \frac{\partial{\psi}}{\partial{P}_{3,3}} 
        = & \: G_{11}{P}_{3,3} + G_{12}({P}_{1,1}+{P}_{2,2}), 
        \\
          \frac{\partial{\psi}}{\partial{P}_{2,1}} 
        = & \: G_{44}({P}_{1,2} + P_{2,1}) + G_{44}^{\prime}({P}_{2,1}-{P}_{1,2}),  
          \quad
          \frac{\partial{\psi}}{\partial{P}_{1,2}} 
        =  G_{44}({P}_{1,2} + P_{2,1}) + G_{44}^{\prime}({P}_{1,2}-{P}_{2,1}), 
        \\ 
           \frac{\partial{\psi}}{\partial{P}_{3,1}} 
        = & \: G_{44}({P}_{1,3} + P_{3,1}) + G_{44}^{\prime}({P}_{3,1}-{P}_{1,3}), 
           \quad
        \frac{\partial{\psi}}{\partial{P}_{1,3}} 
        = G_{44}({P}_{1,3} + P_{3,1}) + G_{44}^{\prime}({P}_{1,3}-{P}_{3,1}), 
        \\ 
        \frac{\partial{\psi}}{\partial{P}_{3,2}} 
        = & \: G_{44}({P}_{2,3} + P_{3,2}) + G_{44}^{\prime}({P}_{3,2}-{P}_{2,3}), 
          \quad
          \frac{\partial{\psi}}{\partial{P}_{2,3}} 
        = G_{44}({P}_{2,3} + P_{3,2}) + G_{44}^{\prime}({P}_{2,3}-{P}_{3,2}).  
     \end{split}
\end{align}

\section*{Appendix C}
\label{sect_appendx_C}
To avoid the huge disparity in the order of parameter values, 
the following units for the material parameters of PbTiO$_3$ are used: 
aJ (attojoule, $10^{-18}$ J), 
nm (nanometer, $10^{-9}$ m), 
aC (attocoulomb, $10^{-18}$ C), 
ps (picosecond, $10^{-12}$ s), 
and V (volt). Accordingly, material parameters in the free energy density 
are \cite{li2002effect} 
\begin{itemize}
        \item {$\alpha_1$ = -0.1725 (aJ)(nm)(aC)$^{-2}$, $\alpha_{11}$ = -0.073 (aJ)(nm)$^5$(aC)$^{-4}$, $\alpha_{12}$ = 0.75 (aJ)(nm)$^5$(aC)$^{-4}$,}
        \item {$\alpha_{111}$ = 0.26 (aJ)(nm)$^9$(aC)$^{-6}$, $\alpha_{112}$ = 0.61 (aJ)(nm)$^9$(aC)$^{-6}$, 
               $\alpha_{123}$ = -3.7 (aJ)(nm)$^9$(aC)$^{-6}$,}
        \item {$c_{11}$ = 174.6 (aJ)(nm)$^{-3}$, $c_{12}$ = 79.37 (aJ)(nm)$^{-3}$, $c_{44}$ = 111.1 (aJ)(nm)$^{-3}$,}
        \item {$q_{11}$ = 11.41 (aJ)(nm)(aC)$^{-2}$, $q_{12}$ = 0.4607 (aJ)(nm)(aC)$^{-2}$, $q_{44}$ = 7.499 (aJ)(nm)(aC)$^{-2}$,}
        \item {$\beta_{11}$ = 0.49586 (aJ)(nm)$^5$(aC)$^{-4}$, $\beta_{12}$ = -0.01459 (aJ)(nm)$^5$(aC)$^{-4}$}
        \item {$G_{11}$ = 0.10 (aJ)(nm)$^3$(aC)$^{-2}$, $G_{12}$ = 0 (aJ)(nm)$^3$(aC)$^{-2}$,
               $G_{44} = G_{44}^{\prime}$ = 0.05 (aJ)(nm)$^3$(aC)$^{-2}$,} 
        \item {$\kappa_c$ = 0.58 (aC)(V)$^{-1}$(nm)$^{-1}$.}
\end{itemize}
For simplicity, the kinetic coefficient related to the domain mobility is assumed to be 
$\zeta = 1$ (aC)$^2$(aJ)$^{-1}$(nm)$^{-1}$(ps)$^{-1}$.
\bibliography{reference}
\end{document}